\documentclass[letterpaper,11pt]{article}
\pdfoutput=1

\usepackage{jheppub}
 
\usepackage{hyperref} 
\usepackage{epsfig}
\usepackage[normalem]{ulem}
\usepackage{bm}
\usepackage{bbm}
\usepackage{slashed}
\usepackage{xspace} 
\usepackage{multirow}

\usepackage{subfig}
\usepackage[countmax]{subfloat}
\newcommand{\setcaptionskip}{\setlength\baselineskip{14pt}}
\newcommand{\setmainskip}{\setlength\baselineskip{18pt}}

\usepackage{amsmath,latexsym}
\usepackage{pstricks}
\usepackage{color} 

\usepackage{arydshln}

\def\cC{\mathcal{C}}

\def\cN{\mathcal{N}}
\def\cS{\mathcal{S}}

\newcommand{\bT}{{\bar T}}

\def\bJ{\mathbf{J}}
\def\bbJ{\mathbf{\bar J}}
\def\bS{\mathbf{S}}
\def\bGamma{\mathbf{\Gamma}}

\newcommand{\eq}[1]{Eq.~\eqref{eq:#1}}
\newcommand{\eqs}[2]{Eqs.~\eqref{eq:#1} and \eqref{eq:#2}} 
\renewcommand{\sec}[1]{Sec.~\ref{sec:#1}}

\newcommand{\fig}[1]{Fig.~\ref{fig:#1}}

\newcommand{\figsthree}[3]{Figs.~\ref{fig:#1}, \ref{fig:#2}, and \ref{fig:#3}}

\newcommand{\img}{\mathrm{i}}

\newcommand{\w}{\omega}

\newcommand{\cM}{\mathcal{M}}
\newcommand{\cO}{\mathcal{O}}

\newcommand{\ddslash}{{d\!\!{}^-}}
\newcommand{\deltaslash}{{\delta\!\!\!{}^-}}

\newcommand{\Ma}{M_{\text{H}_{23}}^{10\oplus\overline{10}}}
\newcommand{\Mb}{M_{\text{H}_{13}}^{10\oplus\overline{10}}}
\newcommand{\Mc}{M_{\text{H}_{12}}^{10\oplus\overline{10}}}
\newcommand{\Maa}{M_{\text{H}_{23}}^{35\oplus\overline{35}}}
\newcommand{\Mbb}{M_{\text{H}_{13}}^{35\oplus\overline{35}}}
\newcommand{\Mcc}{M_{\text{H}_{12}}^{35\oplus\overline{35}}}
\newcommand{\Ha}{\text{H}_{23}}
\newcommand{\Hb}{\text{H}_{13}}
\newcommand{\Hc}{\text{H}_{12}}
\newcommand{\Mdecuplet}{M^{10\oplus\overline{10}}}
\newcommand{\Vdecuplet}{V_g^{10\oplus\overline{10}}}

\newcommand{\IInt}{\int\!\!\!\!\!\int}

\newcommand{\fd}[2]{\parbox{#1}{\includegraphics[width=#1]{#2}}}

\newcommand{\Gma}[1]{\Gamma\left( #1 \right)}

\newcommand{\as}{\alpha_s}

\newcommand{\nn}{\nonumber}

\newcommand{\beq}{\begin{equation}}
\newcommand{\eeq}{\end{equation}}
\newcommand{\bea}{\begin{eqnarray}}
\newcommand{\eea}{\end{eqnarray}}

\newcommand{\Eq}[1]{Equation~\eqref{#1}}
\DeclareRobustCommand{\Sec}[1]{Sec.~\ref{#1}}

\DeclareRobustCommand{\Eq}[1]{Eq.~(\ref{#1})}

\newcommand\prp[1]{\vec{#1}_\perp}
\newcommand\prpsq[1]{\vec{#1}^{\:2}_\perp}
\newcommand\prpsqm[2]{\left( \prp{#1} \!-\! \prp{#2} \right)^2}

\newcommand\prnth[1]{\left(#1\right)}

\newcommand{\bn}{{\bar n}}

\def\bnslash{\bar n\!\!\!\slash}
\def\nslash{n\!\!\!\slash}

\newcommand{\dbar}{d\hspace*{-0.08em}\bar{}\hspace*{0.1em}}
\renewcommand{\tabcolsep}{1ex}

\allowdisplaybreaks[1]

\begin{document}

%%%%%%%%%%%%%%%%%%%%%%%%%%%%%%%%%%%%%%%%%%%%%%%%%%%%%%%%%%%%%%%%%%%%%%%%%%%%%%%%
% Title page
%%%%%%%%%%%%%%%%%%%%%%%%%%%%%%%%%%%%%%%%%%%%%%%%%%%%%%%%%%%%%%%%%%%%%%%%%%%%%%%%

%\date{\today}

\preprint{\vbox{\hbox{MIT-CTP 5808} \hbox{UWThPh 2024-22}} }

\title{\boldmath Reggeization in Color}

\author[1]{Anjie Gao,}
\author[2]{Ian Moult,}
\author[1,3]{Sanjay Raman,}
\author[1]{Gregory Ridgway,}
\author[1,4]{and Iain W. Stewart}

\affiliation[1]{Center for Theoretical Physics, Massachusetts Institute of Technology, Cambridge, MA 02139, USA}
\affiliation[2]{Department of Physics, Yale University, New Haven, CT 06511, USA}
\affiliation[3]{Department of Physics, Harvard University, Cambridge, MA 02138, USA}
\affiliation[4]{University of Vienna, Boltzmanngasse 9, A-1090 Vienna, Austria}

\emailAdd{anjiegao@mit.edu}
\emailAdd{ian.moult@yale.edu}
\emailAdd{sanjayraman@fas.harvard.edu}
\emailAdd{gregridgway@gmail.com}
\emailAdd{iains@mit.edu}

%%%%%%%%%%%%%%%%%%%%%%%%%%%%%%%%%%%%%%%%%%%%%%%%%%%%%%%%%%%%%%%%%%%%%%%%%%%%%%%%
\abstract{
In the high energy limit, $s\gg -t$, amplitudes in planar gauge theories Reggeize, 
with power law behavior $\big( \frac{s}{-t} \big)^{\alpha(t)}$ governed by the 
Regge trajectory $\alpha(t)$.
Beyond the planar limit this simplicity is violated by ``Regge cuts", for which practical 
organizational principles are still being developed.  
We use a top-down effective field theory organization based on color projection in the $t$ channel and
rapidity evolution equations for collinear impact factors, to sum large $s\gg -t$ logarithms for Regge cut contributions.
The results are matrix equations which are closed within a given color channel.  
To illustrate the method we derive  in QCD with $SU(N_c)$ for the first time a closed 6$\times$6 evolution equation for the ``decupletons'' in the $\text{10}\oplus\overline{\text{10}}$ Regge color channel, a 2$\times$2 evolution equation for the ``triantapentons'' in the $\text{35}\oplus\overline{\text{35}}$ color channel, and a scalar evolution equation for the ``tetrahexaconton'' in the 64 color channel.
More broadly, our approach allows us to describe generic Reggeization phenomena in non-planar gauge theories, providing valuable data for the all loop structure of amplitudes beyond the planar limit. 

}
%%%%%%%%%%%%%%%%%%%%%%%%%%%%%%%%%%%%%%%%%%%%%%%%%%%%%%%%%%%%%%%%%%%%%%%%%%%%%%%%

\maketitle
\setmainskip

%\tableofcontents

%%%%%%%%%%%%%%%%%%%%%%%%%%%%%%%%%%%%%%%%%%%%%%%%%%%%%%%%%%%%%%%%%%%%%%%%%%%%%%%%

%%%%%%%%%%%%%%%%%%%%%%%%%%%%%%%%%%%%%%%%%%%%%%%%%%%%%%%%%%%%%%%%
\section{Introduction}
\label{sec:intro}
%%%%%%%%%%%%%%%%%%%%%%%%%%%%%%%%%%%%%%%%%%%%%%%%%%%%%%%%%%%%%%%%

The Regge limit, $|t| \ll s$, is a particularly rich limit of scattering processes in quantum field theory, and has therefore been the focus of intense study since the early days of field theory \cite{Gell-Mann:1964aya,Mandelstam:1965zz,McCoy:1976ff,Grisaru:1974cf,Fadin:1975cb,Kuraev:1976ge,Lipatov:1976zz,Kuraev:1977fs,Balitsky:1978ic,Lipatov:1985uk,Lipatov:1995pn}. From a modern perspective, the Regge limit plays an important role both as a theoretical laboratory for exploring the analytic structure of scattering amplitudes, and in the description of scattering experiments in the small Bjorken-x limit.

The Regge limit is best understood in planar theories. Focusing on the specific case of the $2\to 2$ scattering amplitude in physical kinematics, in the planar limit, one finds the behavior of a pure Regge pole, with terms of the form
\begin{align}
\cM_4 \to \cM_\text{tree} \left( \frac{s}{\mp t} \right)^{\alpha(t)} \,,
\end{align}
governed by the so called Regge trajectory, $\alpha(t)$. For the specific case of the gluon Regge trajectory in QCD we use the notation $\omega_G(t)$. While this form can be shown to hold on general grounds at leading (and next-to-leading) logarithmic order, it can be explicitly derived in specific planar theories where the four point amplitude is known to all loops, such as planar $\cN=4$ super Yang-Mills \cite{Drummond:2007aua}, and its conformal fishnet theory deformation \cite{Korchemsky:2018hnb}.

Beyond the planar limit much less is known about the general all orders structure of the Regge limit. This is due two features which appear beyond the planar limit: the appearance of Regge cuts \cite{Amati:1962nv,Mandelstam:1963cw}, and the proliferation of distinct color representations passed in the $t$-channel. These features begin to appear already at the level of two Reggeons (Glaubers) exchanged in the $t$-channel. In this case, the states can be decomposed into irreducible representations (irreps) as
\begin{align}
  8\otimes8=1\oplus 8^A\oplus 8^S\oplus 10\oplus \overline{10}\oplus 27\oplus 0\,.
\end{align}
With the exception of the $8^A$ and $8^S$, which give rise to pole solutions, the other contributions give rise to cuts. Here the singlet $1$ is the celebrated QCD pomeron \cite{Lipatov:1985uk}, and the other representations were studied in the early literature other representations were studied in the very early literature \cite{Bronzan:1977ag,Bronzan:1977yx,Kwiecinski:1979wr,Praszalowicz:1981zh,Ioffe:2010zz}. Here and below the $0$ denotes an irreducible representation that only exists for $N_c>3$, and for $N_c=3$ this decomposition is the same but with this $0$ deleted.  The case of two-Reggeon (Glauber) exchange is particularly simple since signature \cite{Gribov:1968fc} ensures that the $8^A$ is a pole, and each representation appears with multiplicity one.  However, starting at three-Reggeon (Glauber) exchange, the situation is much more complicated. Decomposing into color irreps, one finds a high multiplicity of identical irreps.  For $N_c>3$ we have
\begin{align}
  8\otimes8\otimes8=1^2 \oplus 8^9 \oplus 10^6 \oplus \overline{10}^6 \oplus 27^6 \oplus 35^2 \oplus \overline{35}^2 \oplus 64 \oplus 0^{17}
  \,,
\end{align}
where the structure of the representations denoted by $0^{17}$ is non-trivial and can be found for example in Ref.~\cite{Keppeler:2012ih}.
For $N_c=3$ this becomes
\begin{align}
  8\otimes8\otimes8=1^2 \oplus 8^8 \oplus 10^4 \oplus \overline{10}^4 \oplus 27^6 \oplus 35^2 \oplus \overline{35}^2 \oplus 64
  \,.
\end{align}
At this level the octet acquires a Regge cut, leading to a breakdown of the naive Regge pole form for the octet amplitude \cite{DelDuca:2001gu}. Understanding the all orders structure of the Regge scattering beyond the planar limit therefore requires the development of new organizational principles.

The problem of Regge cuts, and disentangling Regge cut and Regge pole contributions has been extensively studied in recent years \cite{Bret:2011xm,DelDuca:2011ae,DelDuca:2013ara,Caron-Huot:2013fea,DelDuca:2014cya,Fadin:2016wso,Caron-Huot:2017fxr,Caron-Huot:2017zfo,Fadin:2017nka,Falcioni:2021dgr,Falcioni:2021buo,Falcioni:2020lvv,DelDuca:2021vjq,Caola:2021izf,Falcioni:2021dgr}. Using the Wilson line approach to forward scattering \cite{Caron-Huot:2013fea,Caron-Huot:2017fxr,Caron-Huot:2017zfo}, great successes have been achieved, including an extraction of the three-loop Regge trajectory in QCD \cite{Caola:2021izf,Falcioni:2021dgr}. 

Despite these successes, many questions about the general organization remain. In this paper we focus on the issue of the multiplicity of irreps that appears starting with three-Reggeon (Glauber) exchange. In the planar limit, there is of course also a large multiplicity of irreps, however, they are all degenerate, giving rise to the same Regge asymptotics. To our knowledge, the question of how to organize irreps with high multiplicity has not been considered.  Can they have different Regge asymptotics, split by powers of $1/N_c$? How can one derive closed evolution equations in the case that the irreps have multiplicity bigger than 1?  It is these questions that we address in this paper.

We have recently presented a systematic approach \cite{Gao:2024qsg} to Regge scattering using the effective field theory (EFT) for forward scattering \cite{Rothstein:2016bsq} based on SCET \cite{Bauer:2000ew, Bauer:2000yr, Bauer:2001ct, Bauer:2001yt}. In this approach, the scattering amplitude in the forward limit can be organized in a systematic expansion in the number of Glauber operators exchanged. Each multi-Glauber exchange operator can be factorized into soft and collinear functions. Reggeization is then derived from rapidity renormalization group equations \cite{Chiu:2011qc,Chiu:2012ir} in the EFT, which straightforwardly disentangle cut and pole contributions. We demonstrated some of the power of this approach in \cite{Gao:2024qsg} by reproducing the well known evolution equations for two-Glauber exchange, which could be derived either from collinear impact factors or from a soft function.

In this paper, we extend this approach to three-Glauber exchange. To focus our efforts on the issue of the multiplicity of color representations, we focus on the odderon $1^A$, the ``decupletons'' $10\oplus\overline{10}$, as well as the higher color representations, the $35\oplus\overline{35}$ which we call the ``triantapentons'', and the $64$ which we call the ``tetrahexaconton''.  These two higher color representations only become visible beyond $2\to 2$ scattering. All four cases  
only receive contributions starting from triple Glauber exchange, so we do not have to worry about their mixing with pole contributions and two Glauber contributions. We present a new approach to treat color representations of non-trivial multiplicity in the $t$-channel. Instead of the standard approach of projecting onto irreps using the external particles, we first perform internal color projections to derive evolution equations for different irreps, incorporating their multiplicity. The equations take the form of BFKL equations, but have an additional matrix structure taking into account the multiplicity of the irrep, and the mixing in the internal space. We illustrate this approach by deriving a closed 6$\times$6 evolution equation for the decupleton in QCD with $SU(N_c)$, as well as a 2$\times$2 equation for the triantapenton, and single equations for the odderon and tetrahexaconton.  In a future paper we will combine this with our understanding of the mixing of cut and pole solutions, to address the remaining color channels, most importantly the octet, in our framework.

An outline of this paper is as follows. In \Sec{sec:structure} we describe the organization of the Regge limit in Glauber SCET, highlighting both the organization in terms of the number of Glauber operators, and the organization of color representations.  In \Sec{sec:three_loop} we calculate the renormalization of triple Glauber exchange graphs from the collinear perspective, including both planar and non-planar contributions. By carefully disentangling rapidity divergences, we derive the associated rapidity renormalization group equations. In \Sec{sec:evolution_color} we illustrate our formalism by deriving evolution equations for specific color projections, the odderon, decupletons, triantapentons, and tetrahexaconton, which only receive 
contributions from the three-Glauber graphs. While the evolution equation for the odderon is standard, the evolution equations for the three other color channels are new. We conclude and discuss a number of future directions in \Sec{sec:conc}.

%%%%%%%%%%%%%%%%%%%%%%%%%%%%%%%%%%%%%%%%%%%%%%%%%%%%%%%%%%%%%%%%
\section{Organization of Regge Amplitudes from Glauber SCET}
\label{sec:structure}
%%%%%%%%%%%%%%%%%%%%%%%%%%%%%%%%%%%%%%%%%%%%%%%%%%%%%%%%%%%%%%%%

In this section we describe the general structure of the high energy limit in the organization of the EFT for forward scattering, starting with the factorization and rapidity renormalization results from Ref.~\cite{Gao:2024qsg}. 
We generalize the discussion of $2\to 2$ scattering carried out there by considering the $K\to K$ forward scattering amplitude in the limit where $k$ collimated particles $\{p_{1,1},\ldots,p_{1,k}\}$ travel close to the $n$ direction, and $k'$ collimated particles $\{ p_{2,1},\ldots,p_{2,k'}\}$ travel close to the $\bn$ direction with $k+k'=K$, and forward scatter to particles with momenta $\{p_{4,1},\ldots, p_{4,k}\}$ and $\{p_{3,1},\ldots, p_{3,k'}\}$ respectively. We define the sum of incoming and outgoing momenta in each direction as  $p_j=\sum_i p_{j,i}$. When also specifying whether each of the $k$ particles are quarks, antiquarks, or gluons we refer to the combined state as $\kappa$, and likewise for the $k'$ particles, the combined state is $\kappa'$. There is still only one large rapidity gap between the final state particles $\kappa$ and $\kappa'$, and therefore one type of large logarithm. In particular we let $s = (p_1+ p_2)^2$ and $t = (p_1-p_4)^2$,  so $\ln\big(\frac{s}{-t}\big)$ is a large logarithm with $s\gg -t$. In contrast, $\ln(s_{ij}/s)$ and $\ln(t_{ij}/t)$ are not large logarithms, where we define $s_{ij}=(p_{1,i}+p_{2,j'})^2$ and $t_{ij}=(p_{1,i}- p_{4,j})^2$. This generalization enables us to probe a larger number of color channels.  In this setup the $2\to 2$ scattering is a special case with one particle in each state. Note that we do not allow the emission of mid-rapidity radiation in this setup, since it would lead to a different kinematic situation involving multi-Regge kinematics.

This $K\to K$ scattering amplitude is factorized into collinear impact factors which are jet functions $J$ and $\bar J$ along the $n$-collinear and $\bn$-collinear directions, and a central rapidity soft function $S$. At the bare level we have~\cite{Gao:2024qsg}
\begin{align} \label{eq:factorization_summary}
  \cM_{2\to 2}^{\kappa\kappa'}
  & = \sum_{i=1}^\infty \sum_{j=1}^\infty \, \img \! \IInt_{\perp(i,\,j)}  
  J_{\kappa(i)}^{\{A_i\}}\bigl(\{\ell_{i\perp}\},\epsilon,\eta\bigr)\:
  S_{(i,j)}^{\{A_i\}\,\{A_j'\}}\bigl(\{\ell_{i\perp}\},\{\ell'_{j\perp}\},
   \epsilon,\eta\bigr)\:
  \bar J_{\kappa'(j)}^{\{A'_j\}}\bigl(\{\ell'_{j\perp}\},\epsilon,\eta\bigr)
  \nn\\
  &= i\, \bJ_\kappa(\eta)\cdot \bS(\eta)\cdot \bbJ_{\kappa'}(\eta)
  \,,
\end{align}
where the functions are tied together through transverse momentum integrals,
\begin{align} \label{eq:iint_perp}
  \IInt_{\perp(i,\,j)} 
  \equiv 
  \frac{(-\img)^{i+j}}{i!\, j!}
  \int \prod_{m=1}^{i}\! \frac{\ddslash\!^{d'}\!\ell_{m\perp}}{\vec\ell_{m\perp}^{\,2}} \, \deltaslash^{d'}\!\Bigl(\sum_m \ell_{m\perp}- q_\perp\Bigr) 
  \int\! \prod_{n=1}^{j} \frac{\ddslash\!^{d'}\!\ell'_{n\perp}}{{\vec\ell_{n\perp}^{\,\prime 2}}} \,  \deltaslash^{d'}\!\Bigl(\sum_n {\ell}^{\prime}_{n\perp}- q_\perp\Bigr) \,.
\end{align}
Note that the momentum conservation condition $\sum_m\ell_{m\perp}=\sum_n\ell'_{n\perp}=q_\perp$ 
implies that there are $i-1$ independent $\ell_{m\perp}$'s and $j-1$ independent $\ell'_{n\perp}$'s. 

\begin{figure}
  \begin{center}
  $\fd{7cm}{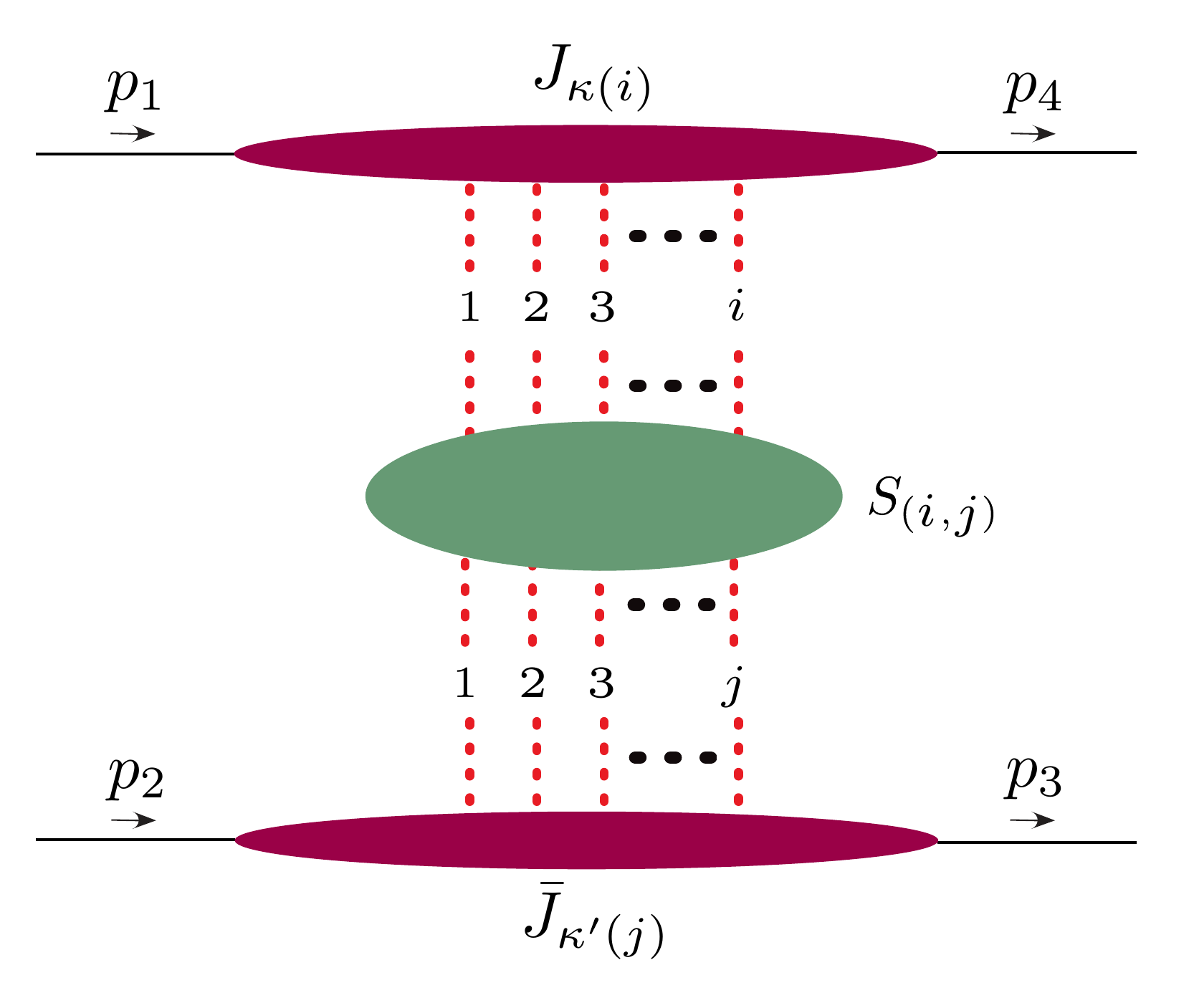} \qquad\qquad\quad\fd{7cm}{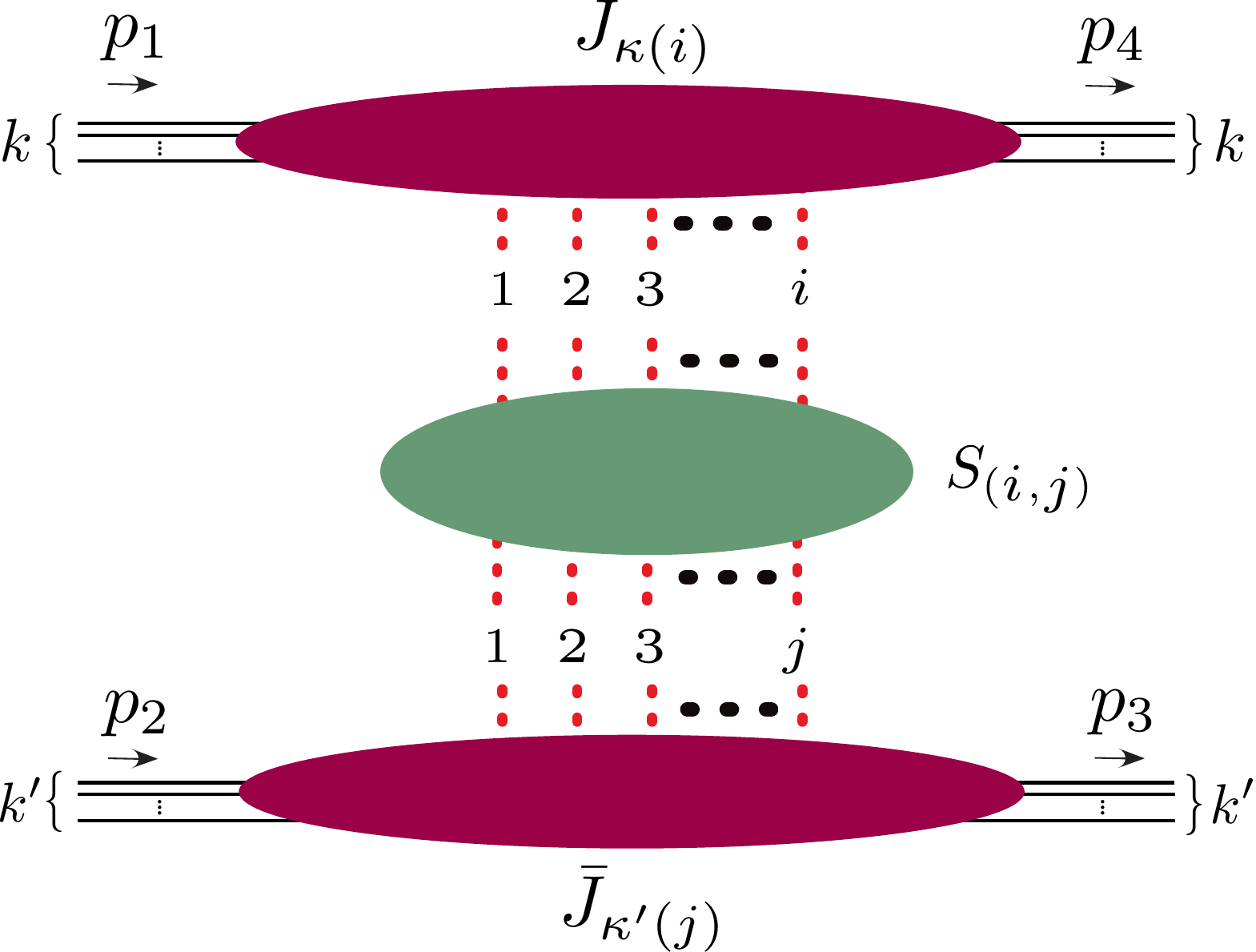}$
  \end{center}
          \vspace{-0.4cm}
          \caption{\setcaptionskip The factorized structure of the $2\to2$ and $K\to K$ forward scattering amplitudes in the Glauber EFT. The amplitude is expressed as a sum over terms with different numbers of Glauber operators exchanges. For a fixed number of Glaubers,  the collinear factors describe the interaction of the projectiles with the Glaubers, while the soft factor describes radiative corrections to the  Glauber potential. }
          \label{fig:fact_diagram}
          \setmainskip
  \end{figure}

In \eq{factorization_summary}, the dimensional regularization parameter $\epsilon=(d-4)/2 = (d'-2)/2$ regulates infrared divergences, while $\eta$ is a rapidity regulator, indicating that the jet and soft functions require rapidity renormalization through counterterms. 
The bare jet function $J_{\kappa(i)}^{A_1 \cdots A_i}(\ell_{1\perp}, \cdots, \ell_{i\perp},\epsilon,\eta)$ can be expressed in terms of the renormalized jet function $J_{\kappa(j)}^{B_1 \cdots B_j}(k_{1\perp},\cdots,k_{j\perp},\epsilon,\nu)$ by 
\begin{align}\label{eq:Jb_Z_Jr}
J_{\kappa(i)}^{A_1 \cdots A_i}(\ell_{1\perp}, \cdots, \ell_{i\perp},\epsilon,\eta)
 & =\, \sum_j \int_{\perp(j)}
  J_{\kappa(j)}^{\{B_j\}}\bigl(\{k_{j\perp}\},\epsilon,\nu\bigr) \:
  Z_{J(j,i)}^{\{B_j\}\, \{A_i\}} \bigl( \{k_{j\perp}\},\{\ell_{i\perp}\},\epsilon,\eta,\nu\bigr) 
  \,, 
\end{align}
where $\int_{\perp(j)}$ is given by
\begin{align}\label{eq:int_perp}
  \int_{\perp(j)} \equiv 
  \frac{(-\img)^{j}}{j!}  \int\bigg[  \prod_{m=1}^{j}\! \frac{\ddslash\!^{d'}\!k_{m\perp}}{\vec k_{m\perp}^{\,2}} \bigg]\, \deltaslash^{d'}  \!\Bigl(\sum_{m=1}^j k_{m\perp}- q_\perp\Bigr) 
 \,.
\end{align}
A shorthand for \eq{Jb_Z_Jr} is $\bJ_\kappa(\eta) = \bJ_\kappa(\nu) \cdot \mathbf{Z_J}$. 
We also have $\bS(\eta)=\mathbf{Z_S} \cdot \bS(\nu) \cdot \mathbf{Z_S}$ and $\bbJ_{\kappa'}(\eta)=\mathbf{Z_J}\cdot \bbJ_{\kappa'}(\nu)$, and SCET soft-collinear consistency implies $\mathbf{Z_S}=\mathbf{Z_J}^{-1}$. 
The rapidity anomalous dimensions for $\mathbf{J}_\kappa(\nu)$ are defined as 
\begin{align}
  \mathbf{\Gamma} \equiv -(\nu \partial_\nu \mathbf{Z_J})\cdot \mathbf{Z_J}^{\!\!\!-1}\,,
\end{align}
where $\partial_\nu = \partial/\partial\nu$, and we will denote the components of $\mathbf{\Gamma}$ as $\gamma_{(i,j)}$.
The resulting rapidity renormalization group equations (RRGEs) are
\begin{align} \label{eq:reggeRRGE}
  \nu\partial_\nu\mathbf{J}_\kappa(\nu)=\mathbf{J}_\kappa(\nu)\cdot\mathbf{\Gamma}\,,
  \qquad
  \nu\partial_\nu\mathbf{S}(\nu)=-\mathbf{\Gamma}\cdot\mathbf{S}(\nu)-\mathbf{S}(\nu)\cdot\mathbf{\Gamma}\,,
  \qquad
  \nu\partial_\nu \mathbf{\bar J}_{\kappa'}(\nu) = \mathbf{\Gamma} \cdot \mathbf{\bar J}_{\kappa'}(\nu)
\,,
\end{align}
and enable a resummation of $\ln\big(\frac{s}{-t}\big)$ logarithms at any order in perturbation theory. It is important to note that the anomalous dimension $\mathbf{\Gamma}$ is independent of the scattering states $\kappa$ and $\kappa'$, even when these are multi-particle states. The only thing that different choices of these external states facilitate is accessing different color channels in the  scattering amplitudes.

The solution of \eq{reggeRRGE} can be written as evolution kernels $\mathbf{U}$ acting on boundary conditions, such as $\bJ_\kappa(\nu) = \bJ_\kappa(\nu_J)\cdot  \mathbf{U}(\nu_J,\nu)$. The jet functions have no large logarithms when evaluated at the rapidity scale $\nu_J=\sqrt{s}$ and the soft function has no large logarithms when evaluated at $\nu_S=\sqrt{-t}$.  Using the RRGE solutions, the amplitude which sums up all large logarithms can therefore be written as 
\begin{align} \label{eq:AmpFACT}
  -\img\cM_{K\to K}^{\kappa\kappa'}=
  \bJ_\kappa(\nu_J)\cdot \mathbf{U}(\nu_J,\nu_S)\cdot \bS(\nu_S)\cdot 
    \mathbf{U}(\nu_S,\nu_J)\cdot \bbJ_{\kappa'}(\nu_J)
  \,.
\end{align}
In general it is difficult to solve for $\mathbf{U}$ in a closed form. One can instead solve for it iteratively
\begin{align} \label{eq:JrgeIt}
  \bJ_\kappa(\nu)&=\bJ_\kappa(\nu_J)+\int_{\nu_J}^{\nu} \frac{d\nu'}{\nu'} \bJ_\kappa(\nu')\cdot\mathbf \Gamma
=\bJ_\kappa(\nu_J)+\int_{\nu_J}^{\nu} \frac{d\nu'}{\nu'} 
\left(\bJ_\kappa(\nu_J)+\int_{\nu_J}^{\nu'} \frac{d\nu''}{\nu''} \bJ_\kappa(\nu'')\cdot\mathbf \Gamma\right)\cdot\mathbf \Gamma
  +\dots \nn
\\
&=\bJ_\kappa(\nu_J)+L_\nu \,\bJ_\kappa(\nu_J)\cdot\mathbf{\Gamma}
+\frac{L_\nu^2}{2!}\bJ_\kappa(\nu_J)\cdot\mathbf{\Gamma}\cdot\mathbf{\Gamma}
+\frac{L_\nu^3}{3!}\bJ_\kappa(\nu_J)\cdot\mathbf{\Gamma}\cdot\mathbf{\Gamma}\cdot\mathbf{\Gamma}
\,+\,\cdots\,.
\end{align}
Here $L_\nu=\ln\frac{\nu}{\nu_J}$, each iteration adds a log, and it is clear that in general the RRGE sums a single logarithmic series.

In the various equations above the dots $\cdot$ are a shorthand which include matrix multiplication in the space of the number of Glaubers $i$, contraction in the space of color indices $\{A_1,\ldots,A_i\}$, and integrations over unconstrained transverse momenta. 
For example, one term in the third order iteration of $\bGamma$'s acting on the leading order jet function looks like
\begin{align}  \label{eq:iteg}
  J_{(2)}^{[0]}\otimes_2\gamma_{(2,3)}\otimes_3\gamma_{(3,3)}\otimes_3\gamma_{(3,4)}=\fd{3cm}{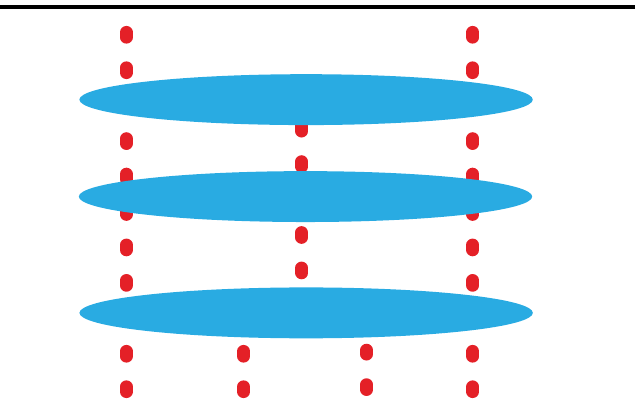}\,,
\end{align}
where $\otimes_j$ has $j-1$ unconstrained transverse momentum integrals.  Here every $\gamma_{(i,j)}$ indicated by a blue blob, is accompanied by a log, and we explicitly see how the number of Glaubers can change during the rapidity evolution. In \eq{iteg} the color index contractions are still suppressed.

As discussed in \cite{Gao:2024qsg}, it is natural to decompose the product of $t$-channel Glauber color octet indices in terms of irreducible representations $R$ of $SU(N_c)$. For a term in the multiplication ($\cdot$) with $m$ Glauber indices $\alpha=A_1\cdots A_m$ we can replace the color index contraction by a sum over irreps,
\begin{align}\label{eq:proj_compl}
  \delta_{m}^{\alpha\alpha'} = \sum_R P_{m\, R}^{\alpha\alpha'} 
   = \sum_R P_{m\, R}^{\alpha\alpha''} P_{m\, R}^{\alpha''\alpha'} \,.
\end{align}
In the sum over $R$, irreps of the same dimension will in general appear multiple times, and do so with unique orthogonal projectors. However, when we have a string of multiplications as in \eq{AmpFACT} or \eq{JrgeIt}, color conservation implies that the objects $\mathbf{U}$, $\mathbf{\Gamma}$, and $\bS$ cannot connect irreps of different dimensions.  (Since irreps of the same dimension appear from different numbers of Glaubers, this does not forbid transitions in the number of Glaubers in neighboring multiplications.)  Writing out the sums over the number of Glaubers and color indices, with implicit sums over repeated indices, use of \eq{proj_compl} allows us to write
\begin{align} \label{eq:AmpFACTcolor}
  -\img\cM_{K\to K}^{\kappa\kappa'}
  &= 
   J^\alpha_{\kappa(i)}(\nu_J)\otimes 
   U_{(i,j)}^{\alpha\beta}(\nu_J,\nu_S)\otimes 
   S_{(j,j')}^{\beta\beta'}(\nu_S)\otimes
   U_{(j',i')}^{\beta'\alpha'}(\nu_S,\nu_J)\otimes 
   \bar J_{\kappa'(i')}^{\alpha'}(\nu_J)
   \\[5pt]
  &=
  \sum_{R,T,T',R'}
   J^{R\alpha}_{\kappa(i)}(\nu_J)\otimes 
   U_{(i,j)}^{RT\alpha\beta}(\nu_J,\nu_S)\otimes 
   S_{(j,j')}^{TT'\beta\beta'}(\nu_S)\otimes
   U_{(j',i')}^{T'R'\beta'\alpha'}(\nu_S,\nu_J)\otimes 
   \bar J_{\kappa'(i')}^{R'\alpha'}(\nu_J)
  \,,\nn
\end{align}
where the sum is constrained such that $\dim(R)=\dim(T)=\dim(T')=\dim(R')$. 

To illustrate better how the equations are used in practical applications, consider the $2\to 2$ scattering amplitude which starts at $\cO(\as)$. Letting $L=\log\frac{s}{-t}$, the summation of terms at order  $\as^{i+1}\sum_n \left(\as L\right)^n$ will be referred to as at N$^i$LL order. The $\as\left(\as L\right)^n$ terms are the leading logs (LL), which resum to the classic LL Regge pole solution for the color octet channel. It can be easily shown that to N$^i$LL, one needs to consider anomalous dimensions with up to $i+1$ Glaubers, such that both the order of anomalous dimensions are number of Glauber exchanges needed are always truncated at a finite level.

At LL and NLL order the sums over irreps in \eq{AmpFACTcolor} are diagonal with $R=T=T'=R'$, since there are unique color representations that do not mix when $i,j,j',i'\le 2$.  In particular, $8\otimes 8 = 1 \oplus 8_A\oplus 8_S \oplus 10 \oplus \overline{10} \oplus 27$, which do not mix with one another (since the two $8$s are distinguished by being of odd and even signature). Furthermore it is well known that resummation for the $10$ and $\overline{10}$ does not occur at NLL. For gluon-gluon scattering the $10\oplus\overline{10}$ has odd signature under $s\to u\simeq -s$, and hence only starts at three Reggeon or three Glauber exchange. In our formalism we see this explicitly by checking that the color of the leading order jet function $J_{g(2)}^{[0]}$ (which is the same as $C_H$ in \eq{CH} below) projects to zero for this decupleton.  Our focus in this paper will be to consider the leading cases where the sums over irreps become non-diagonal, which begins for amplitudes at NNLL order, which involves computing graphs that start at 3-loop order.

We also remark that while the internal objects in \eq{AmpFACTcolor} involve matrix color structures within irreps of a common dimension, the combinations that the total scattering amplitude is sensitive to depend on what we are scattering in the states $\kappa\kappa'$. 
In particular, if we let $\lambda$ denote the color indices of the state $\kappa$ and we again use \eq{proj_compl}, we can write
\begin{align}
  J_{\kappa \lambda (i)}^{R\alpha} = J_{\kappa \lambda'(i)}^{R\alpha} \delta^{\lambda'\lambda}
   = \sum_U J_{\kappa \lambda'(i)}^{R\alpha} P_{i\: U}^{\lambda'\lambda}
   \equiv \sum_U J_{\kappa U\lambda(i)}^{R\alpha} 
  \,.
\end{align}
Here $J_{\kappa U\gamma(i)}^{R\alpha}$ is a transition (collinear) impact factor between the color irrep $U$ for the scattering particles $\kappa$, and the color irrep $R$ for the $i$ attached Glaubers. 
For one particle scattering states we again have a $t$-channel projection from the one-particle in and out states that leads to unique color irreps. For example when we scatter a gluon,  $\kappa=g$, the in-gluon with color index $C$ and out-gluon with color index $D$ can be decomposed as an $8\otimes 8$, where the product $\lambda=CD$ can be written as a sum of irreps that can not mix.  Therefore, in this case the upper projectile color state $U$ provides a unique projection onto a linear combination of the Glauber color states $R$ with the same dimension, $\dim(U)=\dim(R)$.   In our formalism the internal matrix structure will be revealed by anomalous dimension calculations which only depend on the Glaubers and can be determined independent of the scattering states.  However to have full access to the elements of these matrices will require scattering more than one particle in the states $\kappa$ and $\kappa'$.

%%%%%%%%%%%%%%%%%%%%%%%%%%%%%%%
\begin{figure}%[ht]
  \centering
\subfloat[]{\label{fig:J1g1S1J1}  
  \includegraphics[scale=0.3]{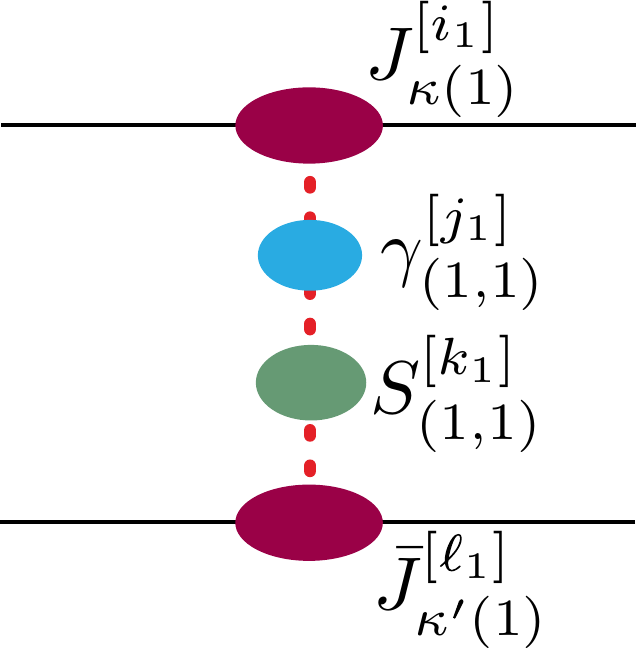} }\hspace{2cm}
  \subfloat[]{\label{fig:J2g2S2J2}
  \includegraphics[scale=0.3]{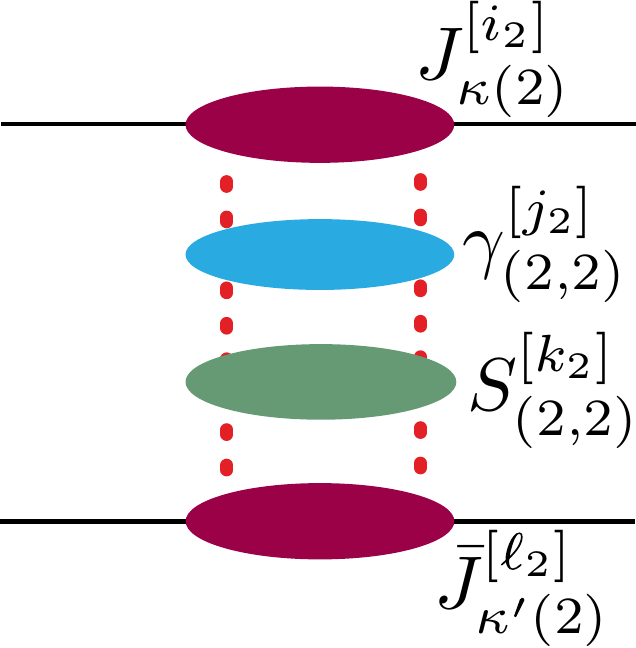} }
  \\
    \subfloat[]{\label{fig:J3g3S3J3}  
  \includegraphics[scale=0.3]{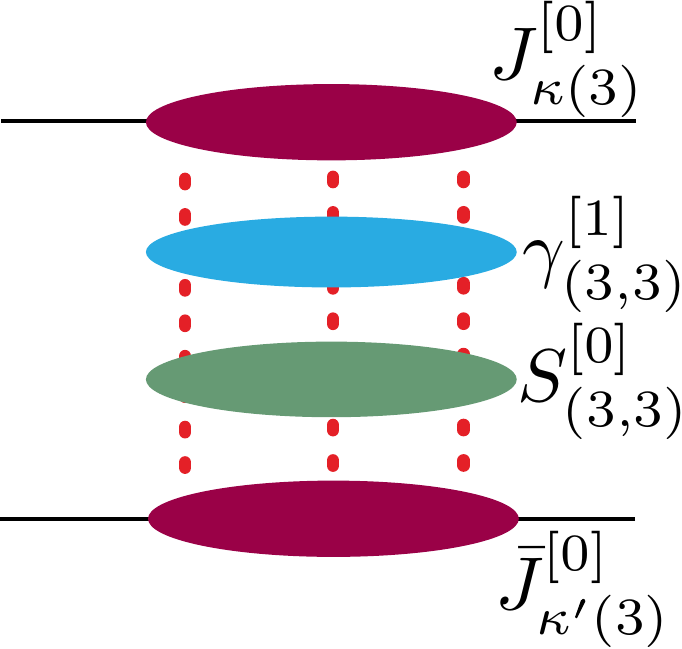} }\hspace{1cm}
  \subfloat[]{\label{fig:J2g23S3J3}  
  \includegraphics[scale=0.3]{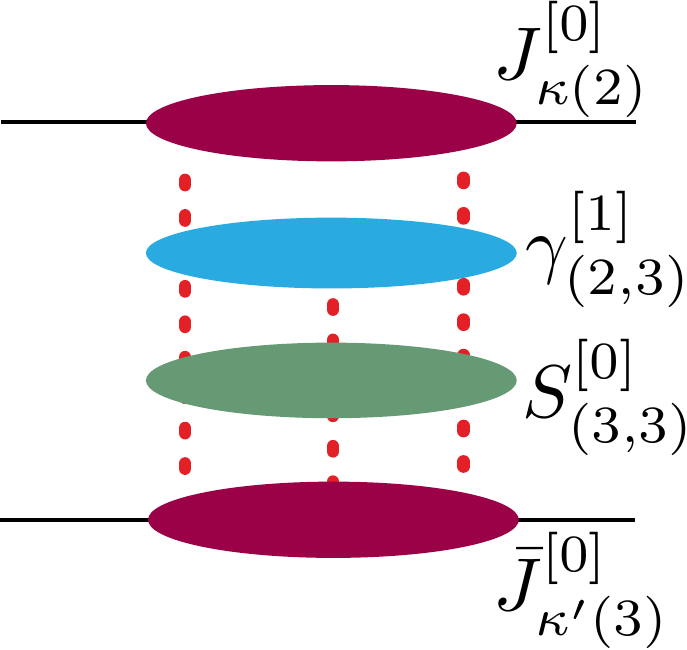} }\hspace{1cm}
  \subfloat[]{\label{fig:J3g32S2J2}  
  \includegraphics[scale=0.3]{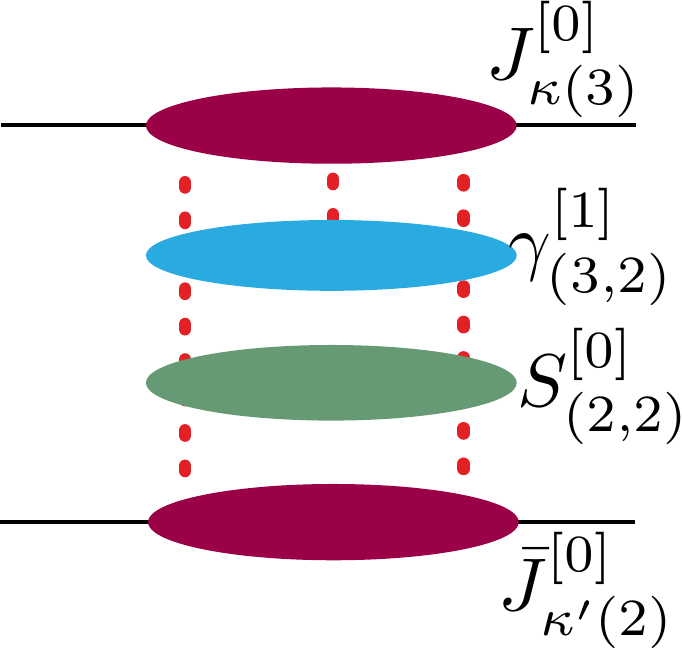} }
  \caption{All the EFT contributions with one logarithm (associated with $\gamma_{(i,j)}$, the blue blob) up to three loops. Here we have omitted the contribution where $\gamma$ is put between $S$ and $\bar J$.
  This is also a complete set of ingredients needed for resummation up to NNLL. In our notation, jet and soft functions start at tree level, with $[0]$ on the superscript, while $\gamma_{(i,j)}$ starts with at least one loop, i.e., $j_1,\,j_2\ge1$.
  (a) One-Glauber-exchange contribution. $\gamma_{(1,1)}$ here is our all-order definition of the gluon Regge trajectory. At one loop, one needs to consider $\gamma_{(1,1)}$ at one loop and everything else at tree level, i.e. $j_1=1$, $i_1=k_1=\ell_1=0$. At two or three loops, one needs to consider contributions with $i_1+j_1+k_1+\ell_1=2$ or $3$.
  (b) Two-Glauber-exchange contribution. Since two Glaubers give a loop and $\gamma_{(2,2)}$ is at least one loop, this contribution starts at two loops, with $j_2=1$ and $i_2=k_2=\ell_2=0$. At three loops, we have $i_2+j_2+k_2+\ell_2=2$.
  (c, d, e) Three-Glauber-exchange contribution. Since three Glaubers give two loops and $\gamma$ is at least one loop, this contribution starts at three loops, where the jet and soft functions are all at tree level.
  }
  \label{fig:1log_3loop}
\end{figure}
%%%%%%%%%%%%%%%%%%%%%%%%%%%%%%%

\fig{1log_3loop} lists all the contributions in our EFT to three loops with one large logarithm, which are thus at NNLL order 
(the trivial symmetric contributions obtained from switching $n$ with $\bar n$ are not shown).
The one large log is associated with a single anomalous dimension in the blue blob labeled $\gamma$, while the jet and soft functions there should be interpreted as in their canonical rapidity scales $\nu_J$ and $\nu_S$ respectively, so that they do not contain large rapidity logarithms.
The jet functions, soft functions and anomalous dimensions displayed in \fig{1log_3loop} are the complete set of EFT ingredients needed for NNLL resummation. Here the superscripts in square brackets, $[k]$, indicate that the term is needed at $k$-loop order. 
At LL one only needs to consider the one-Glauber-exchange contribution in \fig{J1g1S1J1}, with the one-loop Regge trajectory $\gamma_{(1,1)}^{[1]}$  resummed to all order (which means we add to the diagram by iterating it along its attached vertical lines).
At NLL, one needs to include contributions from the two-loop Regge trajectory $\gamma_{(1,1)}^{[2]}$ and one-loop correction to $J_{\kappa(1)}(\nu_J)\, S_{(1,1)}(\nu_S)\, \bar J_{\kappa'(1)}(\nu_J)$ in \fig{J1g1S1J1}, and starting from \fig{J2g2S2J2} iterations of the one-loop $\gamma_{(2,2)}^{[1]}$.
At NNLL, one should include the three-loop Regge trajectory $\gamma_{(1,1)}^{[3]}$ and two-loop corrections to $J_{\kappa(1)}(\nu_J)\, S_{(1,1)}(\nu_S)\, \bar J_{\kappa'(1)}(\nu_J)$ in \fig{J1g1S1J1}, two-loop $\gamma_{(2,2)}^{[2]}$ and one-loop correction $J_{\kappa(2)}^{(1)}$, $S_{(2,2)}^{(1)}$, $\bar J_{\kappa'(2)}^{(1)}$ in \fig{J2g2S2J2}, as well as $\gamma_{(3,3)}^{[1]}$, $\gamma_{(2,3)}^{[1]}$, $\gamma_{(3,2)}^{[1]}$ in \figsthree{J3g3S3J3}{J2g23S3J3}{J3g32S2J2}. Note that by symmetry $\gamma_{(2,3)}$ and $\gamma_{(3,2)}$ are identical.

Before proceeding to our specific analysis, some general remarks are in order.  In the framework presented above  $\gamma_{(2,2)}^{[1]}$ was calculated in \cite{Gao:2024qsg}, reproducing the traditional BFKL Pomeron, $8^S$, $27$. The two-Glauber exchange contribution includes $8^A$ channel as well: this part can be thought as the unitarization of the LL Regge pole solution, 
which occurs from higher order terms in the field theory in the Glauber exchange field theory (in contrast to being manifest from the start in the Reggeon exchange language). 
In traditional studies of Regge amplitudes, organized in terms of Reggeon exchange, there is no separation of the collinear constants in our collinear impact factors $J_{\kappa(i)}(\nu_J)$ from the soft constants that appear in our $S_{(i,j)}(\nu_S)$.
Instead, the so-called impact factor $C_{\kappa}$ is interpreted as the loop corrections to the projectile, and is extracted by subtracting the logarithmic terms from the Regge limit of the QCD amplitudes. These $C_{\kappa}$ are known to two loops for $\kappa=q$ and $\kappa=g$~\cite{Fadin:1993wh,Fadin:1994fj,Fadin:1993qb,DelDuca:1998kx,Caola:2021izf,Falcioni:2021dgr}.
The impact factors for one-Reggeon exchange (color octet) should therefore be identified as $J_{\kappa(1)}(\nu_J)\sqrt{S_{(1,1)}(\nu_S)}$.  However, this identification can be tricky starting at two-loops (NNLL), since the definition of the impact factors can be a ``scheme choice'' depending on how one separates the ``pole'' contribution from the ``cut'' contributions~\cite{Falcioni:2021dgr}.  The color channels that we focus on here do not exhibit this mixing, and hence we will not explain how this issue is resolved in the Glauber EFT here, and instead leave this discussion to future work.

In this vein, we can also outline what other contributions that occur at the three Glauber level are left for discussion in a future paper.
The three Glauber contributions include signature even channels for the Pomeron, $8^S$, and $27$, which provide terms required by unitarization in the these channels, which we will not discuss here.  More interestingly, the three-loop Regge trajectory was extracted recently~\cite{Caola:2021izf,Falcioni:2021dgr} from the full QCD amplitudes by subtracting ``Regge cut'' contributions using the Wilson line approach to define Reggeon exchange amplitudes. In the Wilson line approach, two organizational schemes were given~\cite{Falcioni:2021buo}, leading to two results for the three-loop Regge trajectory. In the ``multi-Reggeon scheme'' the contributions from multiple Reggeon states are separated out, and the three loop gluon Regge trajectory is defined by contributions connected to the single Reggeon state.  In the alternative ``cut scheme'' the Regge cut contribution is defined to be the non-planar part of the multi-Reggeon contribution, so that planar parts are grouped into the three loop gluon trajectory. In the cut scheme consistent universal results for the Regge trajectory were found for $gg$, $qg$ and $qq$ scattering~\cite{Caola:2021izf,Falcioni:2021dgr,Caola:2022dfa}. 
It is therefore interesting to see which scheme is favored by the organization in the Glauber exchange picture, which we also leave to future work. 

In the following section we begin with a calculation of the anomalous dimensions $\gamma_{(3,3)}^{[1]}$ and $\gamma_{(2,3)}^{[1]}$ that first occur from three Glauber exchange, and which are necessary to access the information on the color channels that we study here. Then in \sec{evolution_color} we carry out the color projections and present evolution equations for the odderon ($1^A$), decupleton ($10\oplus\overline{10}$), triantapenton ($35\oplus\overline{35}$), and tetrahexaconton ($64$).

%%%%%%%%%%%%%%%%%%%%%%%%%%%%%%%%%%%%%%%%%%%%%%%%%%%%%%%%%%%%%%%%
\section{Rapidity Renormalization of Triple-Glauber Exchange}
\label{sec:three_loop}
%%%%%%%%%%%%%%%%%%%%%%%%%%%%%%%%%%%%%%%%%%%%%%%%%%%%%%%%%%%%%%%%

In this section, we derive the leading order rapidity RGE for the contribution to the forward scattering amplitude mediated by three Glauber operators,
$J_{\kappa(3)}$. We will refer to this as the triple-Glauber exchange contribution. From our general discussion above, it takes the form
\begin{align}\label{eq:RGE_J3}
  \nu\partial_\nu J_{\kappa(3)}=
   &  J_{\kappa(2)} \otimes \gamma_{(2,3)}
   +  J_{\kappa(3)} \otimes \gamma_{(3,3)}\,.
\end{align}
Recall that $J_{\kappa(i)}$ starts at $\cO(g^i)$,  $\gamma_{(3,3)}$ starts at $\cO(g^2)$, and $\gamma_{(3,2)}$ starts at $\cO(g^3)$, so that both terms in \eq{RGE_J3} are equally important.
Since our results will be independent of the choice of the scattering state $\kappa$, we can work with single particle states $\kappa=g,q$ at intermediate steps. 

The tree-level contribution of the three-Glauber-exchange to forward scattering was calculated in \cite{Rothstein:2016bsq}, and gives
\begin{align}\label{eq:double_box}
  -\img\cM_{(3,3)}^{[0]} &=\fd{3cm}{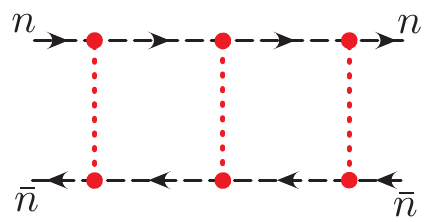}
   = J_{(3)}^{[0]}\otimes S_{(3,3)}^{[0]}\otimes \bar J_{(3)}^{[0]}
   \\[.2cm]& \nn
   = 2 \prnth{T^{A_1} T^{A_2}T^{A_3}} \otimes \prnth{\bT^{A_1} \bT^{A_2} \bT^{A_3}} {\cal S}^{n\bn}\:g^6 \:  
  \frac{1}{3!}\,
  \int\! \frac{\ddslash\!^{d'}\ell_{1\perp}\ddslash\!^{d'}\ell_{2\perp}}
   {\vec \ell_{1\perp}^{\;2}\, \vec \ell_{2\perp}^{\;2}\, \bigl(\vec q_\perp-\vec\ell_{1\perp}-\vec\ell_{2\perp}\bigr)^{2} } 
   \,,
\end{align}
where $T^A$ are color generators for the $n$-collinear state $\kappa$ and $\bar T^A$ are color generators for the $\bn$-collinear state $\kappa'$, and ${\cal S}^{n\bn}$ contains factors that track the external particle spin. For example for $\kappa\kappa'=q\bar q$ we have
\begin{align} \label{eq:Snbn}
  \cS^{n\bn} = \Big[ \bar u_n  \frac{\bnslash}{2} u_n \Big]\Big[ \bar v_\bn  \frac{\nslash}{2} v_\bn \Big] \,.
\end{align} 

To determine the three Glauber exchange anomalous dimensions, we find it easiest to perform the calculation from the collinear perspective developed in \cite{Gao:2024qsg}, where we compute $n$-collinear loops to three-Glauber-exchange graphs. To one-loop order, this is interpreted as the one-loop $J_{(3)}$ convolved with the tree level $S_{(3,3)}$ and $\bar J_{(3)}$,
\begin{align}
\sum -\img\cM_{(3,3)}^{\text{one $n$ loop}} = J_{(3)}^{[1]}\otimes S_{(3,3)}^{[0]}\otimes \bar J_{(3)}^{[0]}\,.
\end{align}
By comparing the rapidity poles in this result with the tree level result, we are then able to extract the anomalous dimensions $\gamma_{(2,3)}$ and $ \gamma_{(3,3)}$ in \eq{RGE_J3}. As is standard, when performing this comparison, one must keep the transverse integrals of $\ell_{1\perp}$, $\ell_{2\perp}$ and $k_\perp$ undone, to properly renormalize the contribution from the tree level triple-Glauber exchange. 

The anomalous dimension we derive does not depend on the external states, and therefore we consider $n$-collinear quark $\bn$-collinear antiquark scattering for simplicity.
In \sec{graphs}, we compute all the rapidity divergent one-loop collinear graphs for triple-Glauber exchange. In \sec{RRGE}, we read off the rapidity anomalous dimensions by comparing the rapidity poles for these graphs in \sec{graphs} with the double box diagram.

%%%%%%%%%%%%%%%%%%%%%%%%%%%%%%%%%%%%%%%%%%%%%%%%%%%%%%%%%%%%%%%%
\subsection{Calculation of Rapidity Divergent Graphs}
\label{sec:graphs}
%%%%%%%%%%%%%%%%%%%%%%%%%%%%%%%%%%%%%%%%%%%%%%%%%%%%%%%%%%%%%%%%

%%%%%%%%%%%%%%%%%%%%%%%%%%%%%%%
\begin{figure}[t!]
  \begin{center}
  \hspace{-0.55cm}
  $-\img\cM_{\rm P}
  \equiv\fd{3cm}{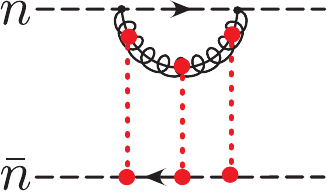}  \hspace{10.8cm}
$
  \\[15pt]
  \hspace{-0.55cm}
  $-\img\cM_{\rm{NP_1}}\equiv  \fd{3cm}{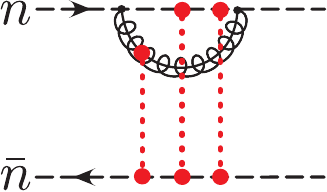}
  $
  \hspace{0.3cm}
$-\img\cM_{\rm{NP_2}}\equiv \fd{3cm}{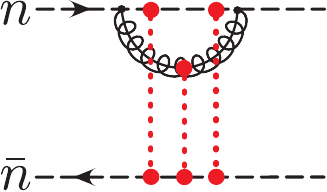}$  \hspace{0.3cm}
$-\img\cM_{\rm{NP_3}}\equiv\fd{3cm}{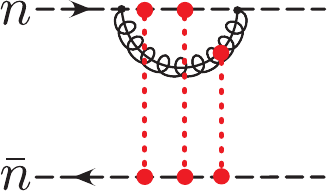}$  
\\[15pt]
  \hspace{-0.55cm}
  $-\img\cM_{\rm{NP_4}}\equiv  \fd{3cm}{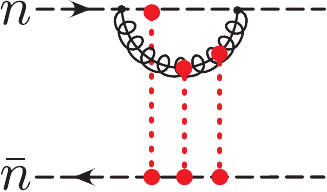}$
  \hspace{0.3cm}
$-\img\cM_{\rm{NP_5}}\equiv \fd{3cm}{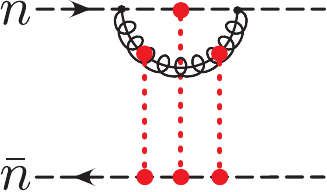}$  \hspace{0.3cm}
$-\img\cM_{\rm{NP_6}}\equiv\fd{3cm}{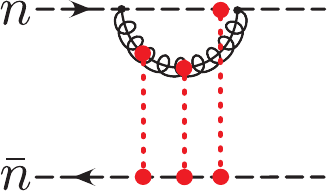}$
  \end{center}
    \vspace{-0.4cm}
    \caption{\setcaptionskip The one planar and six non-planar one-loop collinear diagrams that contribute to the rapidity renormalization group evolution of the three-Glauber state.
    The notation ``P'' and ``NP'' denotes planar and non-planar.
    Notice that although we consider quark-antiquark scattering graphs, the $\gamma_{(3,3)}$ and $\gamma_{(2,3)}$ are universal.
    }
    \label{fig:diagrams_1c_loop_3Glaubers}
    \setmainskip
  \end{figure}
%%%%%%%%%%%%%%%%%%%%%%%%%%%%%%%

All the rapidity divergent graphs for quark-antiquark scattering with one collinear loop and three Glauber exchanges are shown in \fig{diagrams_1c_loop_3Glaubers}. These include one planar and six non-planar graphs. Since the details of the calculations are similar to \cite{Gao:2024qsg}, we simply present the final results for the required $1/\eta$ rapidity divergent terms.

%%%%%%%%%%%%%%%%%%%%%%%%%%%%%%%
\subsubsection{Planar Graph}
%%%%%%%%%%%%%%%%%%%%%%%%%%%%%%%

For the single planar graph, we find
\begin{align}
 & -\img\cM_{\text{P}}
 = \frac{4\as (4\pi \as)^3 w^2}{3!\,\eta} \biggl(\frac{\sqrt{s}}{\nu} \biggr)^{-\eta} 
  P_\text{P} 
	\int\!\! \frac{\dbar^{d'}\!\ell_{1\perp} \dbar^{d'}\!\ell_{2\perp}}{\vec\ell_{1\perp}^{\,2} \vec\ell_{2\perp}^{\,2} \vec\ell_{3\perp}^{\,2} }
  \int\!\! \frac{\dbar^{d'}\!k_\perp \; \vec q_\perp^{\,2}}{
	\prpsq{k} \prpsqm{k}{q}
} + \cO(\eta^0)
\,,
\end{align}
where  $\ell_{3\perp}=q_\perp-\ell_{1\perp}-\ell_{2\perp}$ and $P_{\text{P}}$ captures all color and spinor factors
\begin{align}
  P_\text{P} &= \prnth{-\img f^{A_1B_1C}}\prnth{-\img f^{A_2CB_2}} \prnth{-\img f^{A_3CB_2}} \prnth{T^{B_1} T^{B_2}} \otimes \prnth{\bT^{A_1} \bT^{A_2} \bT^{A_3}} \cS^{n\bn}\,.
\end{align}
Here $B_i$'s are adjoint color indices attaching to the $n$-quark line, $A_i$'s are adjoint color indices attaching to the $\bn$-antiquark line, and $\cS^{n\bn}$ is the spinor factor in \Eq{eq:Snbn}.
%

%%%%%%%%%%%%%%%%%%%%%%%%%%%%%%%%%%%%%%%%%%%%%%%%%%%%%%%%%%%%%%%%
\subsubsection{Non-Planar Graphs}
%%%%%%%%%%%%%%%%%%%%%%%%%%%%%%%%%%%%%%%%%%%%%%%%%%%%%%%%%%%%%%%%

For the six non-planar graphs we find
\begin{align}
  -\img\cM_{{\rm NP_1}}&=\frac{4\as (4\pi \as)^3 w^2}{3!\,\eta} \left(\frac{\sqrt{s}}{\nu} \right)^{-\eta}   P_{\rm NP_1}
	\int\!\! \frac{\dbar^{d'}\!\ell_{1\perp} \dbar^{d'}\!\ell_{2\perp}}{\vec\ell_{1\perp}^{\,2} \vec\ell_{2\perp}^{\,2} \vec\ell_{3\perp}^{\,2}}
  \int\!\! \frac{\dbar^{d'}\!k_\perp \; \vec \ell_{1\perp}^{\,2}}{
	\prpsq{k} \bigl(\vec\ell_{1\perp}-\vec k_\perp\bigr)^2
} + \cO(\eta^0)
\,,\\ \nn
-\img\cM_{{\rm NP_2}}&=\frac{4\as (4\pi \as)^3 w^2}{3!\,\eta} \left(\frac{\sqrt{s}}{\nu} \right)^{-\eta}   P_{\rm NP_2}
	\int\!\! \frac{\dbar^{d'}\!\ell_{1\perp} \dbar^{d'}\!\ell_{2\perp}}{\vec\ell_{1\perp}^{\,2} \vec\ell_{2\perp}^{\,2} \vec\ell_{3\perp}^{\,2}}
  \int\!\! \frac{\dbar^{d'}\!k_\perp \; \vec \ell_{2\perp}^{\,2}}{
	\prpsq{k} \bigl(\vec\ell_{2\perp}-\vec k_\perp\bigr)^2
} + \cO(\eta^0)
\,,\\ \nn
-\img\cM_{{\rm NP_3}}&=\frac{4\as (4\pi \as)^3 w^2}{3!\,\eta}  \left(\frac{\sqrt{s}}{\nu} \right)^{-\eta}   P_{\rm NP_3}
	\int\!\! \frac{\dbar^{d'}\!\ell_{1\perp} \dbar^{d'}\!\ell_{2\perp}}{\vec\ell_{1\perp}^{\,2} \vec\ell_{2\perp}^{\,2} \vec\ell_{3\perp}^{\,2}}
  \int\!\! \frac{\dbar^{d'}\!k_\perp \; \vec \ell_{3\perp}^{\,2}}{
	\prpsq{k} \bigl(\vec\ell_{3\perp}-\vec k_\perp\bigr)^2
} + \cO(\eta^0)
\,,\\ \nn
-\img\cM_{{\rm NP_4}}&=\frac{4\as (4\pi \as)^3 w^2}{3!\,\eta} \left(\frac{\sqrt{s}}{\nu} \right)^{-\eta}   P_{\rm NP_4}
	\int\!\! \frac{\dbar^{d'}\!\ell_{1\perp} \dbar^{d'}\!\ell_{2\perp}}{\vec\ell_{1\perp}^{\,2} \vec\ell_{2\perp}^{\,2} \vec\ell_{3\perp}^{\,2}}
  \int\!\! \frac{\dbar^{d'}\!k_\perp \; \bigl(\vec \ell_{2\perp}+\vec \ell_{3\perp}\bigr)^2}{
	\prpsq{k} \bigl(\vec\ell_{2\perp}+\vec\ell_{3\perp}-\vec k_\perp\bigr)^2
} + \cO(\eta^0)
\,,\\ \nn 
-\img\cM_{{\rm NP_5}}&=\frac{4\as (4\pi \as)^3 w^2}{3!\,\eta} \left(\frac{\sqrt{s}}{\nu} \right)^{-\eta}   P_{\rm NP_5}
	\int\!\! \frac{\dbar^{d'}\!\ell_{1\perp} \dbar^{d'}\!\ell_{2\perp}}{\vec\ell_{1\perp}^{\,2} \vec\ell_{2\perp}^{\,2} \vec\ell_{3\perp}^{\,2}}
  \int\!\! \frac{\dbar^{d'}\!k_\perp \; \bigl(\vec \ell_{1\perp}+\vec \ell_{3\perp}\bigr)^2}{
	\prpsq{k} \bigl(\vec\ell_{1\perp}+\vec\ell_{3\perp}-\vec k_\perp\bigr)^2
} + \cO(\eta^0)
\,,\\ \nn
-\img\cM_{{\rm NP_6}}&=\frac{4\as (4\pi \as)^3 w^2}{3!\,\eta} \left(\frac{\sqrt{s}}{\nu} \right)^{-\eta}   P_{\rm NP_6}
	\int\!\! \frac{\dbar^{d'}\!\ell_{1\perp} \dbar^{d'}\!\ell_{2\perp}}{\vec\ell_{1\perp}^{\,2} \vec\ell_{2\perp}^{\,2} \vec\ell_{3\perp}^{\,2}}
  \int\!\! \frac{\dbar^{d'}\!k_\perp \; \bigl(\vec \ell_{1\perp}+\vec \ell_{2\perp}\bigr)^2}{
	\prpsq{k} \bigl(\vec\ell_{1\perp}+\vec\ell_{2\perp}-\vec k_\perp\bigr)^2
} + \cO(\eta^0)
\,,
\end{align}
where again $\ell_{3\perp}=q_\perp-\ell_{1\perp}-\ell_{2\perp}$.
The color and spinor factors are contained in the prefactors $P_{\text{NP}_i}$. We write these as a color factor $C_{\text{NP}_i}$, multiplying the spinor factor $\cS^{n\bn}$ of \Eq{eq:Snbn},
\begin{align}
  P_{\text{NP}_i} &= C_{\text{NP}_i}\, \cS^{n\bn}\,.
\end{align}
The explicit color factors are given by
\begin{align}
  C_{\text{NP}_1} &= \prnth{-\img f^{A_1B_1B_2}} \prnth{T^{B_1} T^{A_2} T^{A_3} T^{B_2}} \otimes \prnth{\bT^{A_1} \bT^{A_2} \bT^{A_3}} \,,
  \\ \nn
  C_{\text{NP}_2} &= \prnth{-\img f^{A_2B_1B_2}} \prnth{T^{B_1} T^{A_1} T^{A_3} T^{B_2}} \otimes \prnth{\bT^{A_1} \bT^{A_2} \bT^{A_3}} \,,
  \\ \nn
  C_{\text{NP}_3} &= \prnth{-\img f^{A_3B_1B_2}} \prnth{T^{B_1} T^{A_1} T^{A_2} T^{B_2}} \otimes \prnth{\bT^{A_1} \bT^{A_2} \bT^{A_3}} \,,
  \\ \nn
  C_{\text{NP}_4} &= \prnth{-\img f^{A_2B_1C}} \prnth{-\img f^{A_3CB_2}} \prnth{T^{B_1} T^{A_1} T^{B_2}} \otimes \prnth{\bT^{A_1} \bT^{A_2} \bT^{A_3}} \,,
  \\ \nn
  C_{\text{NP}_5} &= \prnth{-\img f^{A_1B_1C}} \prnth{-\img f^{A_3CB_2}} \prnth{T^{B_1} T^{A_2} T^{B_2}} \otimes \prnth{\bT^{A_1} \bT^{A_2} \bT^{A_3}} \,,
  \\ \nn
  C_{\text{NP}_6} &= \prnth{-\img f^{A_1B_1C}} \prnth{-\img f^{A_2CB_2}} \prnth{T^{B_1} T^{A_3} T^{B_2}} \otimes \prnth{\bT^{A_1} \bT^{A_2} \bT^{A_3}} \,.
\end{align}

%%%%%%%%%%%%%%%%%%%%%%%%%%%%%%%
\subsection{Rapidity Renormalization Group Equations}
\label{sec:RRGE}
%%%%%%%%%%%%%%%%%%%%%%%%%%%%%%%

We can now derive the rapidity RGE for $J_{(3)}$. The procedure for the rapidity renormalization of multi-Glauber exchange graphs from the collinear perspective was developed in \cite{Gao:2024qsg}. In doing so, it is essential to match not only the momentum structure of the integrals, but also the color structure, with the tree level triple-Glauber graph. To do so, we must ``unravel" the color in the above rapidity divergent graphs. This can be achieved using the identity  $[T^A,T^B]=\img f^{ABC}T^C$. Graphically, we have 
\begin{align}
  \cC:\; \fd{1.5cm}{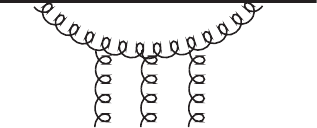} &=  \fd{1.5cm}{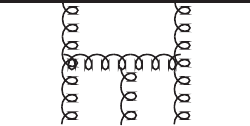} 
    \,,\\[.1in]
  \cC:\; \fd{1.5cm}{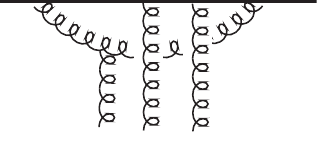} &=  \fd{1.5cm}{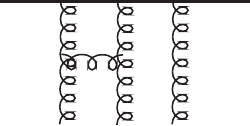} + \fd{1.5cm}{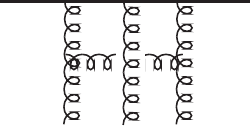} 
  - \frac{N_c}{2}\, \fd{1.5cm}{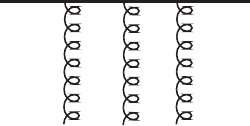}
    \,,\nn\\[.1in]
  \cC:\; \fd{1.5cm}{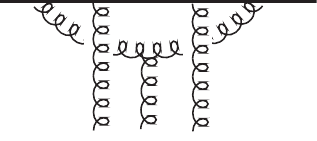} &=  \fd{1.5cm}{figsNNLL/C_TC.pdf} + \fd{1.5cm}{figsNNLL/C_r12.pdf} +
  \fd{1.5cm}{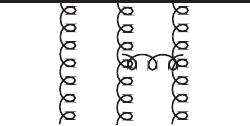} 
  - \frac{N_c}{2}\, \fd{1.5cm}{figsNNLL/C_tree.pdf}
    \,,\nn\\[.1in]
  \cC:\; \fd{1.5cm}{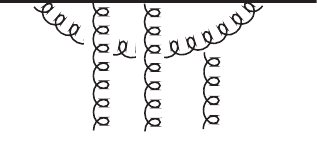} &=  \fd{1.5cm}{figsNNLL/C_r23.pdf} + \fd{1.5cm}{figsNNLL/C_r13.pdf} 
  - \frac{N_c}{2}\, \fd{1.5cm}{figsNNLL/C_tree.pdf}
    \,,\nn\\[.1in]
  \cC:\; \fd{1.5cm}{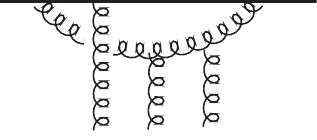} &=  \fd{1.5cm}{figsNNLL/C_r23.pdf}  
  - \frac{N_c}{2}\, \fd{1.5cm}{figsNNLL/C_tree.pdf}
    \,,\nn \\[.1in]
  \cC:\; \fd{1.5cm}{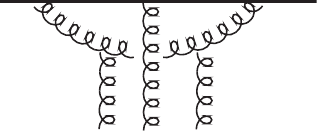} &=  \fd{1.5cm}{figsNNLL/C_r13.pdf}  
    \,,\nn \\[.1in]
  \cC:\; \fd{1.5cm}{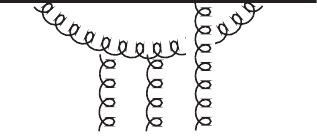} &=  \fd{1.5cm}{figsNNLL/C_r12.pdf}  
  - \frac{N_c}{2}\, \fd{1.5cm}{figsNNLL/C_tree.pdf}
    \,,\nn
\end{align}
where $\cC$ indicates that these diagrams should be interpreted in terms of color flow, not momentum flow. Ultimately, we decompose the color structures of the collinear graphs to ``fundamental'' base objects, by which we mean that there are only ``horizontal'' and ``vertical'' lines. 

Once organized in this manner, we can then directly read off the rapidity RGE for $J_{\kappa(3)}$, which is given by
\begin{align} \label{eq:collRRGE}
 &  \nu\frac{\partial}{\partial \nu} J_{\kappa(3)}^{\alpha}(\{\ell_{\perp}\};\nu) \\
 & =\, -\int\! \prod_{i=1}^{3}\frac{\dbar^{d'}\!k_{i\perp}}{k_{i\perp}^2} \,
   J_{\kappa(3)}^{\beta}(\{k_{\perp}\};\nu)
   \biggl\{ \prod_{i=1}^3\delta^{A_iB_i}\, 
   \delta^{d'}\!\bigl(\ell_{i\perp} \!-\! k_{i\perp}\bigr)    
   \sum_{i=1}^3\w_G(\ell_{i\perp}) 
  \nn\\
 &\quad + 
    \delta^{d'}\!\bigl(q_\perp\!-\!k_{1\perp}\!-\!k_{2\perp}\!-\!k_{3\perp}\bigr)\,
    \frac2{N_c} \sum_{i<j}^3 \delta^{d'}\bigl(\ell_{i\perp}\!+\!\ell_{j\perp}\!-\!k_{i\perp}\!-\!k_{j\perp}\bigr)\, 
    C_{{\rm H}_{ij}}^{\beta\alpha}\, K_\text{NF}(\ell_{i\perp}, \ell_{j\perp};k_{i\perp},k_{j\perp})  \biggr\}
  \nn\\
  & +\int\!\prod_{i=1}^{2}\frac{\dbar^{d'}\!k_{i\perp}}{k_{i\perp}^2} \,
    \delta^{d'}\!(q_\perp-k_{1\perp}-k_{2\perp})
    J_{\kappa(2)}^{B_1B_2}(k_{1\perp},k_{2\perp};\nu)\,
    C_{\rm TC}^{B_1B_2\,A_1A_2A_3}\,K_{\rm TC}(k_{1\perp},k_{2\perp};\ell_{1\perp},\ell_{2\perp},\ell_{3\perp})
  .\nn
 \end{align}
Here we have used $\ell_{\perp}$ and $k_{\perp}$ in the argument of $J_{(3)}$ to denote $\{\ell_{1\perp},\,\ell_{2\perp},\,\ell_{3\perp}\}$ and $\{k_{1\perp},\,k_{2\perp},\,k_{3\perp}\}$, and $\alpha=A_1A_2A_3$ and $\beta=B_1B_2B_3$ are shorthand notations for color indices appearing in $J_{(3)}$ and $C_{{\rm H}_{ij}}$.
The color factors $C_{{\rm H}_{ij}}$ and $C_{\rm TC}$ are defined as
 \begin{align}\label{eq:CHij}
  &C_{{\rm H}_{12}}^{\beta\alpha}= -(- \img f^{A_1 B_1 C} )(-\img f^{A_2 B_2 C}) \delta^{A_3B_3} 
  \,,
  \qquad
  C_{{\rm H}_{23}}^{\beta\alpha}= -(- \img f^{A_2 B_2 C} )(-\img f^{A_3 B_3 C}) \delta^{A_1B_1} 
  \,,
  \nn\\
  &C_{{\rm H}_{13}}^{\beta\alpha}= -(- \img f^{A_1 B_1 C} )(-\img f^{A_3 B_3 C}) \delta^{A_2B_2}
  \,, 
  \qquad
  C_{\rm TC}^{B_1B_2\,A_1A_2A_3}= \img
  f^{A_1 B_1 C} f^{A_2 C D} f^{A_3 B_2 D} \,,
 \end{align}
and 
\begin{align}
  & \omega_G(\ell_\perp) = -\as N_c \int \frac{\dbar^{d'}\!k_\perp \; \vec\ell_\perp^{\,2}}{\prpsq{k} \prpsqm{\ell}{k}} \,,
  \\
  & K_\text{NF}(\ell_{i\perp},\ell_{j\perp}; k_{i\perp},k_{j\perp}) = 
  \as N_c
  \prnth{-\frac{\bigl(\vec \ell_{i\perp}+\vec \ell_{j\perp}\bigr)^{\!2}}{\vec k_{i\perp}^2\,\vec k_{j\perp}^2} 
  + \frac{\vec\ell_{i\perp}^{\,2}}{\vec k_{i\perp}^2 \bigl(\vec \ell_{i\perp} - \vec k_{i\perp}\bigr)^2} 
  + \frac{\vec \ell_{j\perp}^{\,2}}{\vec k_{j\perp}^2 \bigl(\vec \ell_{j\perp} - \vec k_{j\perp}\bigr)^2} } \,,
  \nn \\
  &
  K_{\rm TC}(\{k_{\perp}\};\{\ell_{\perp}\}) 
  =
  \frac{\bigl(\vec\ell_{1\perp}+\vec \ell_{2\perp}\bigr)^2}{\vec k_{1\perp}^{\,2}\bigl(\vec\ell_{3\perp}-\vec k_{2\perp}\bigr)^2}
  +\frac{\bigl(\vec\ell_{2\perp}+\vec\ell_{3\perp}\bigr)^2}{k_{2\perp}^2\bigl(\vec\ell_{1\perp}-\vec k_{1\perp}\bigr)^2}
  -\frac{\ell_{2\perp}^2}{\bigl(\vec\ell_{1\perp}-\vec k_{1\perp}\bigr)^2\bigl(\vec\ell_{3\perp}-\vec k_{2\perp}\bigr)^2}
  -\frac{\vec q_\perp^2}{\vec k_{1\perp}^2 \vec k_{2\perp}^2}
  ,\nn
\end{align}
where we have used $\{k_{\perp}\}$ and $\{\ell_{\perp}\}$ in the argument to denote $\{k_{1\perp},\,k_{2\perp}\}$ and $\{\ell_{1\perp},\,\ell_{2\perp},\,\ell_{3\perp}\}$.
It is interesting to note that our $K_{\rm TC}$ from $2\to 3$ Glauber exchange coincides with the kernel extracted from the $1\to3$ transition of Reggeons in the Wilson approach~\cite{Caron-Huot:2017fxr}.

Graphically, we have
\begin{align}
  \nu\frac{\partial}{\partial\nu}\fd{3.5cm}{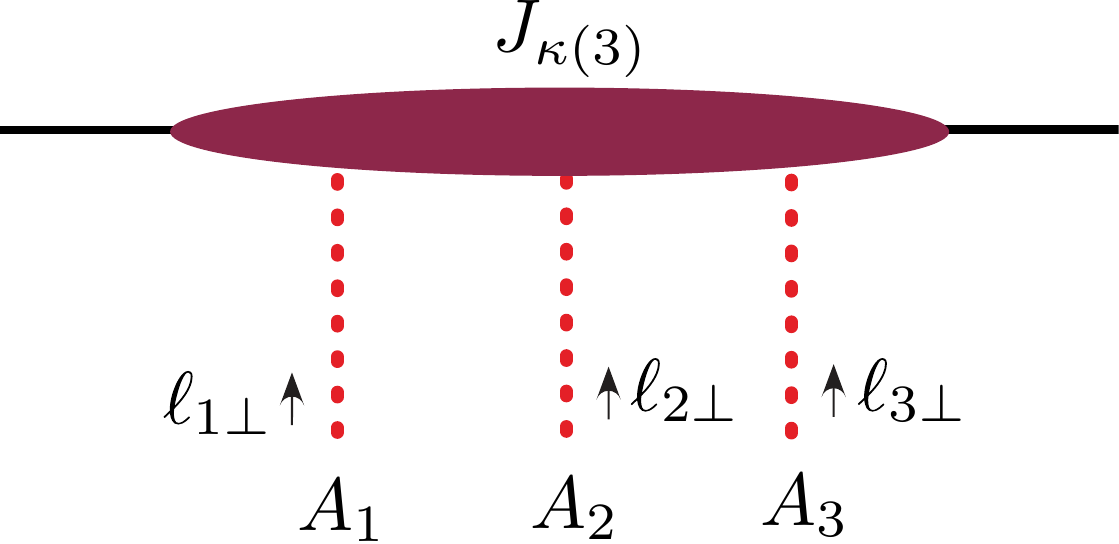} &= \fd{3.5cm}{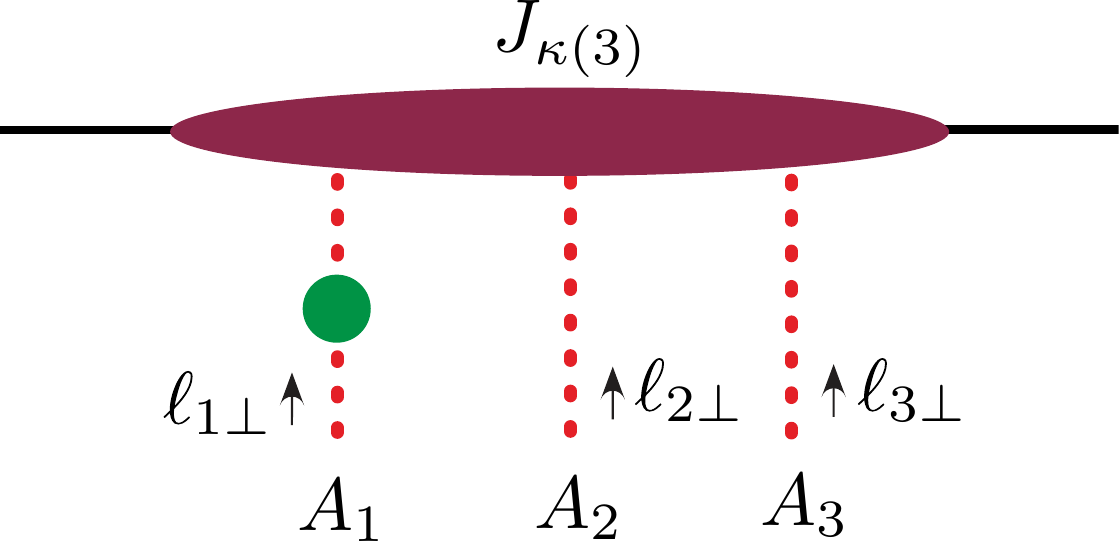}+\fd{3.5cm}{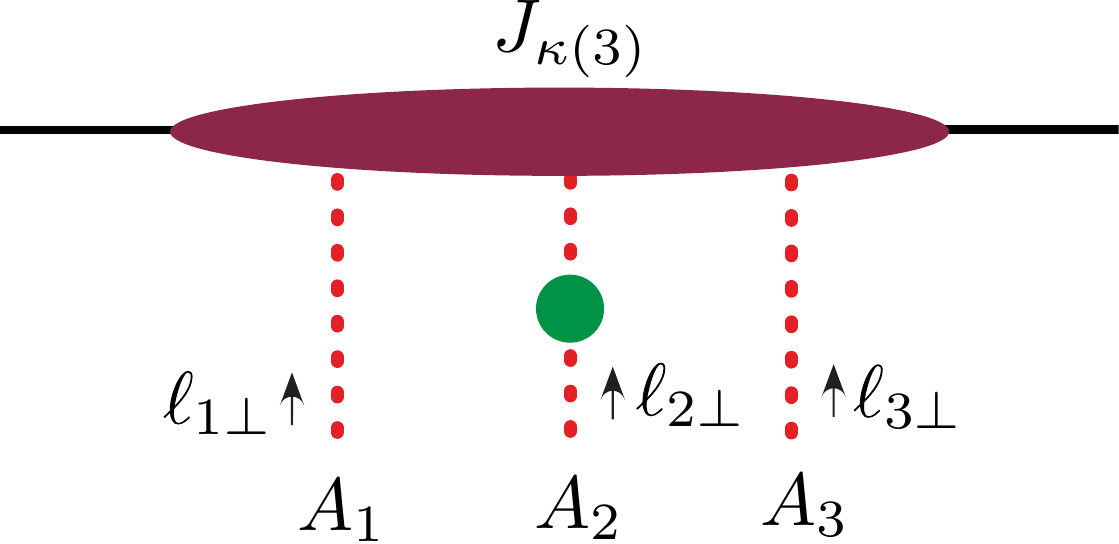}+\fd{3.5cm}{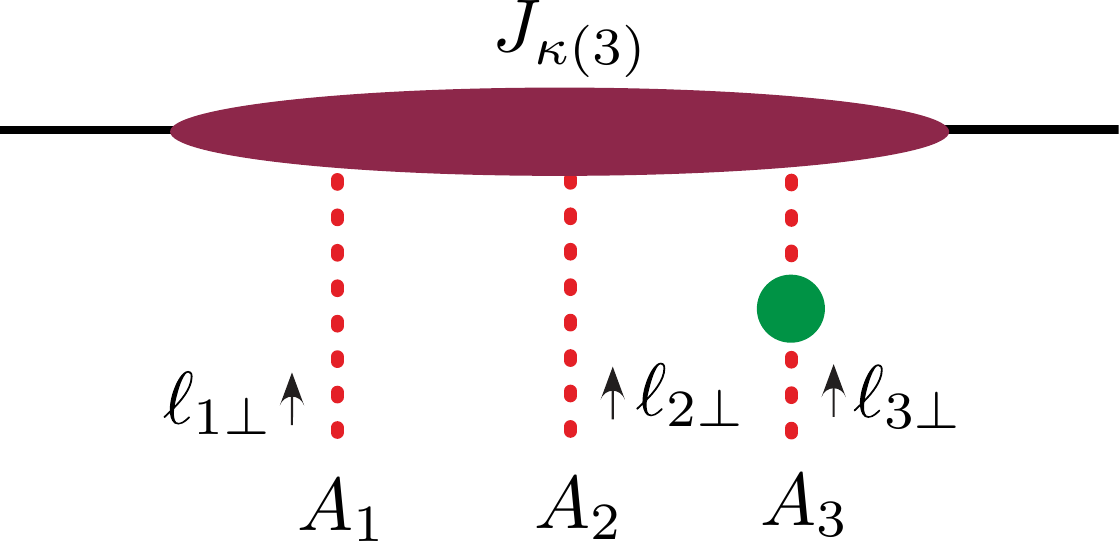}
  \nn\\&
  +\fd{3.5cm}{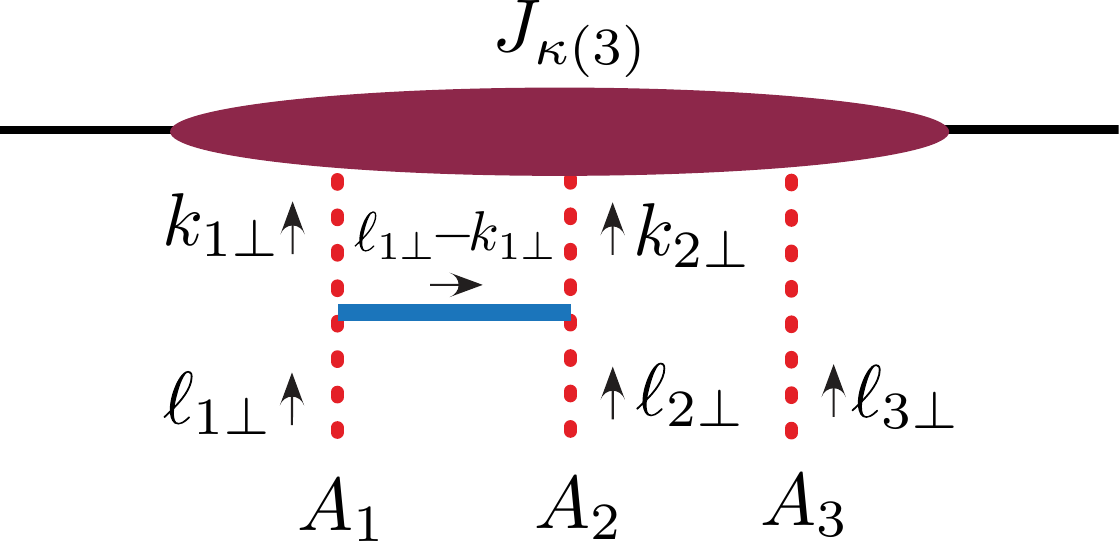}+\fd{3.5cm}{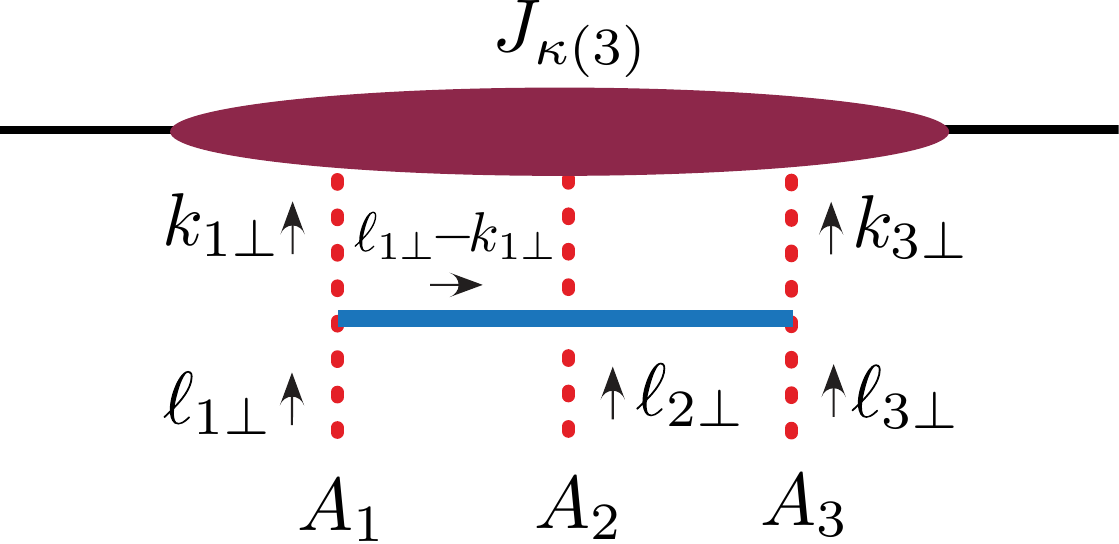}+\fd{3.5cm}{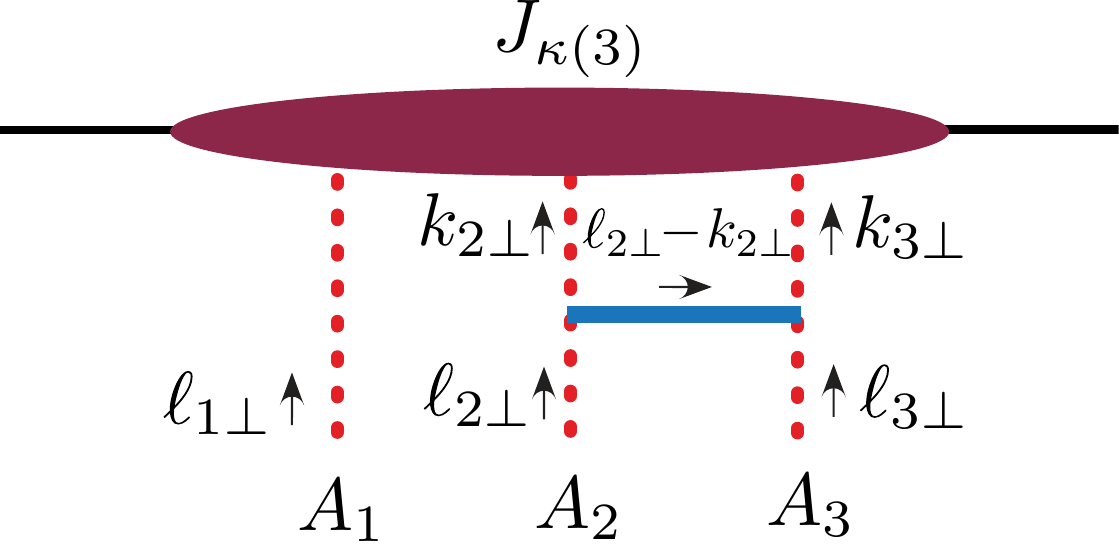}
  \nn\\&
  +\fd{3.5cm}{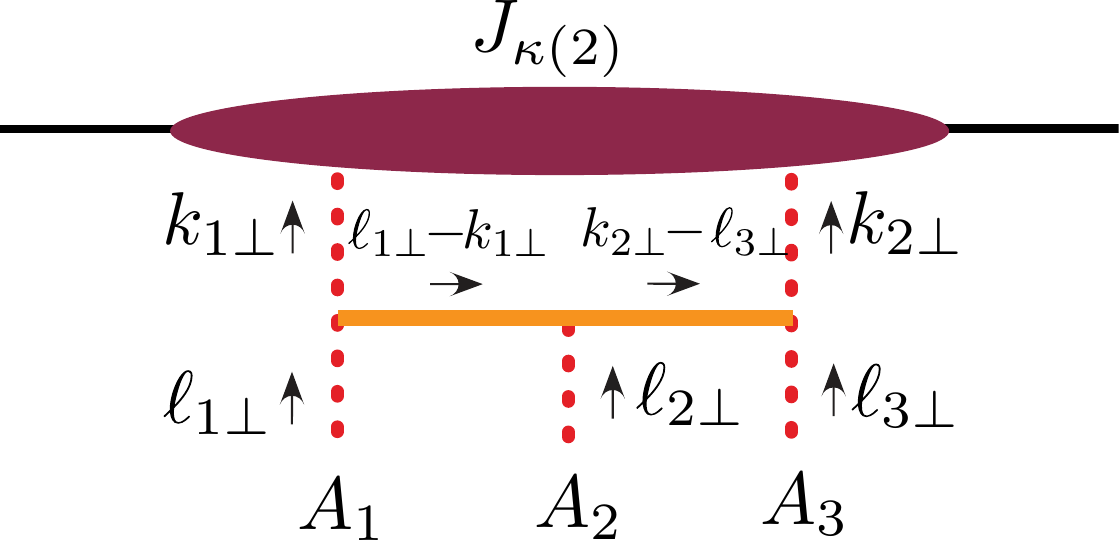}\nn
\end{align}
where the green blobs denote $\w_G$'s, the blue lines denote $K_{\rm NF}$'s, and the orange line denotes $K_{\rm TC}$. We can write $\gamma_{(3,3)}$ and $\gamma_{(2,3)}$ graphically as
\begin{align}\label{eq:gamma33_graph}
  \fd{3.2cm}{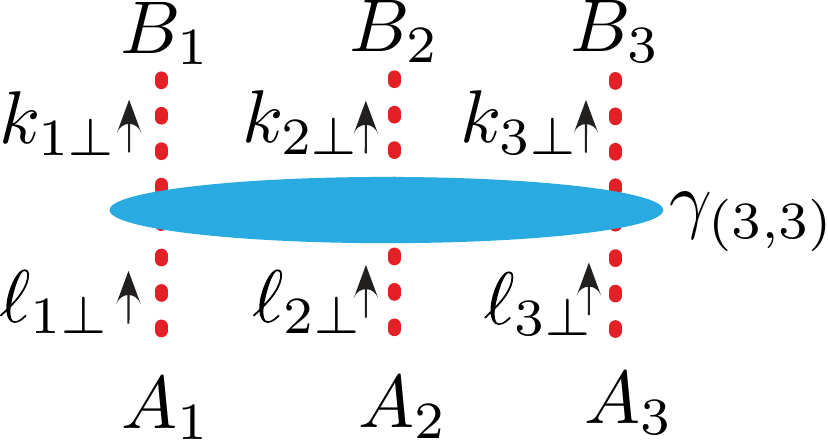}\, &= \fd{3.2cm}{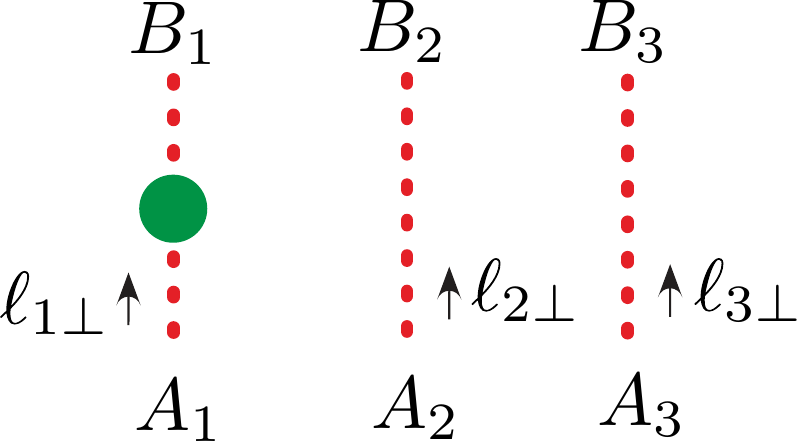}+\fd{3.2cm}{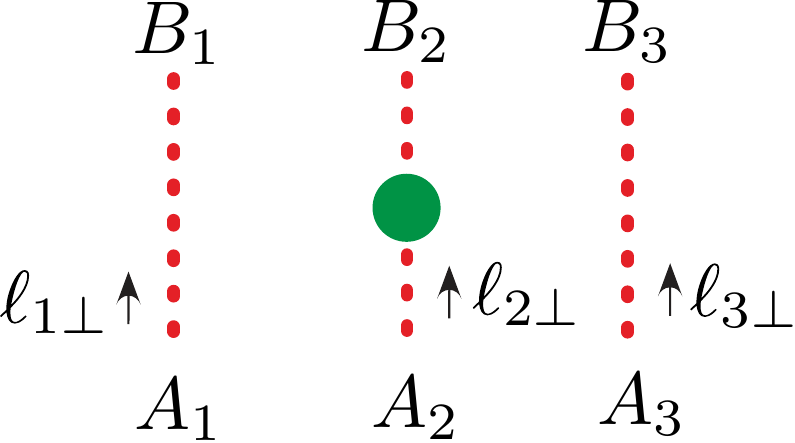}+\fd{3.2cm}{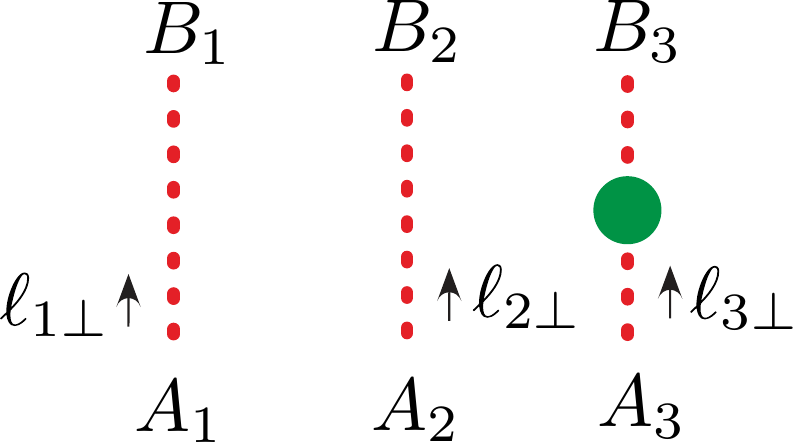}
  \\[.2cm]&
  +\fd{3.2cm}{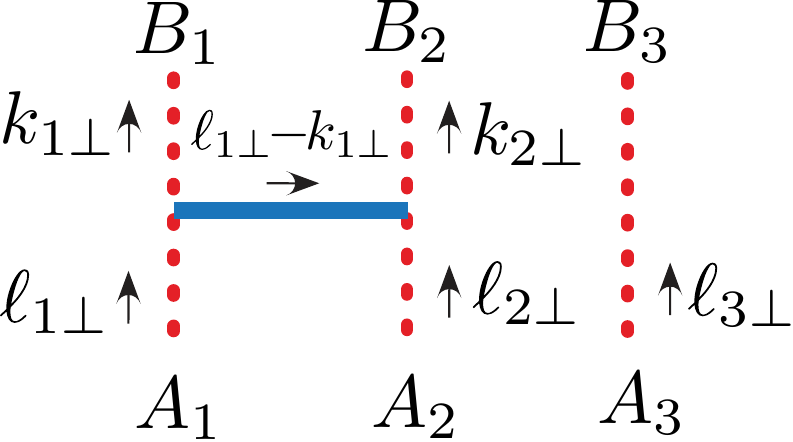}+\fd{3.2cm}{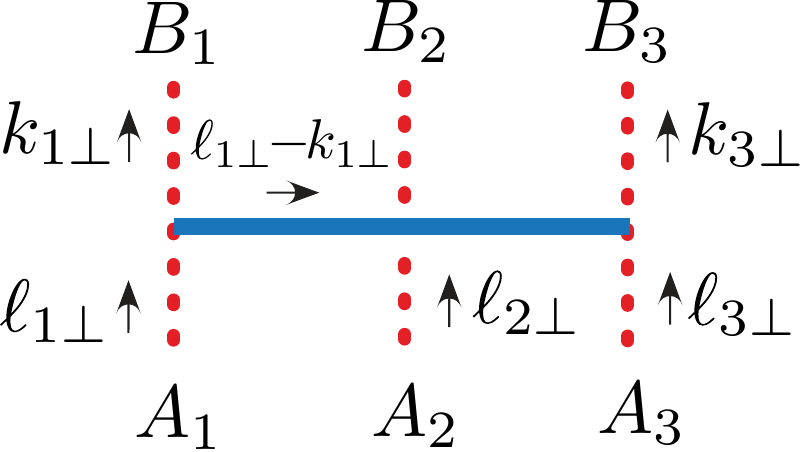}+\fd{3.2cm}{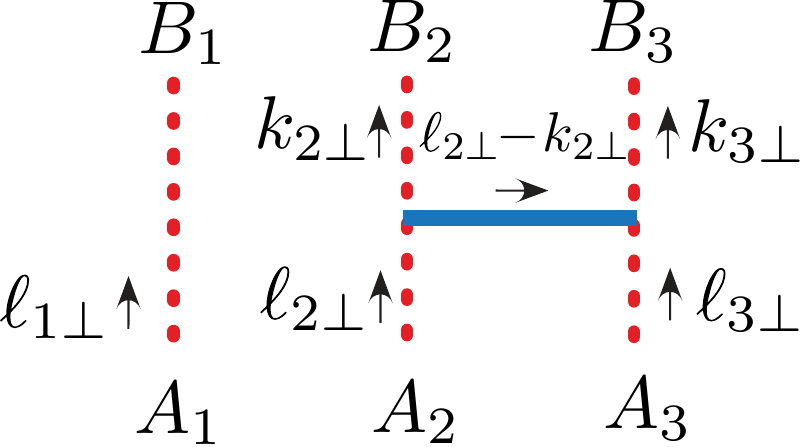}
  \,,
  \nn\\[.5cm]
  \fd{3.2cm}{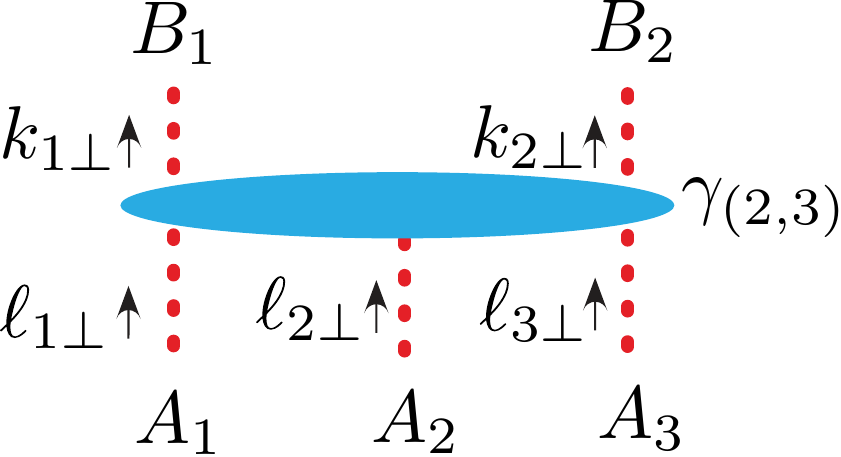}\, &= \fd{3.2cm}{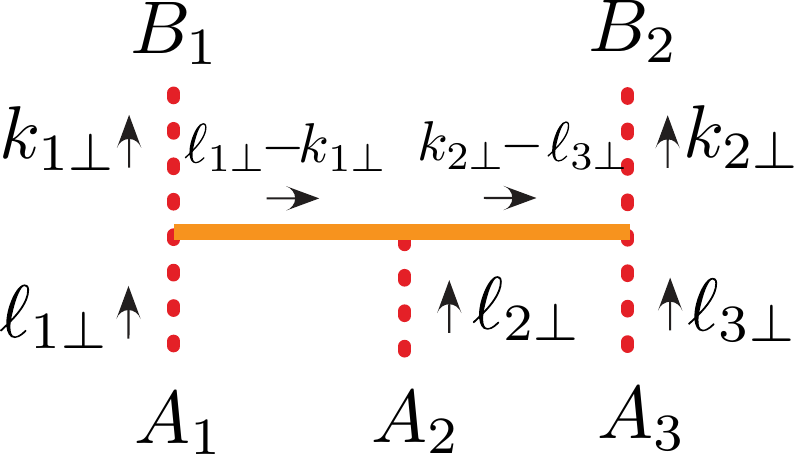}
  \,.
\end{align}

%%%%%%%%%%%%%%%%%%%%%%%%%%%%%%%%%%%%%%%%%%%%%%%%%%%%%%%%%%%%%%%%
\section{Color Evolution Equations for Triple-Glauber Exchange}
\label{sec:evolution_color}
%%%%%%%%%%%%%%%%%%%%%%%%%%%%%%%%%%%%%%%%%%%%%%%%%%%%%%%%%%%%%%%%

The rapidity renormalization group equation in Eq.~\eqref{eq:collRRGE} contains open color indices. To proceed, we need to understand how to project it down onto t-channel irreps in the case that the irreps have non-trivial multiplicity. This will allow us to study, for example, different $10\oplus \overline{10}$ irreps to understand if they can have different Regge asymptotics. Our approach will provide a new way of organizing color in the Regge limit. Instead of doing color projections in terms of the external particles, we propose to first perform internal color projections to derive RRGEs for different color channels. This allows us to perform the resummation for any specific channel, and then finally do specific projections of this resummed result onto the external particles in the scattering process.

For the case of two Glauber exchange the internal and external color projection approaches are equivalent, since after doing either color projection, the RRGEs are scalar in color.
This occurs because the decomposition of two-Glauber color indices into irreducible representations (irreps) is relatively simple, and the irreps are all distinguished by either dimension or parity
\begin{align}\label{eq:2gluon_decomp}
  8\otimes8=1\oplus 8^A\oplus 8^S\oplus 10\oplus \overline{10}\oplus 27\,.
\end{align}
Importantly, there is only one copy for each channel $R$, 
i.e., the multiplicity for each channel here is 1. 
For example, $\gamma_{(2,2)}^{[1]}$ has two color structures, each of which corresponds to {\it a number} for any color channal once we do a color projection onto channel $R$,
\begin{align}
  \fd{1.2cm}{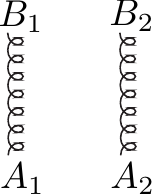} &= C_{\delta}  \equiv  \delta^{A_1 B_1} \delta^{A_2 B_2} = \sum_R P_{2 \, R}^{A_1 A_2\, B_1 B_2}\,,\\[.2cm]
  \label{eq:CH}
  \fd{1.2cm}{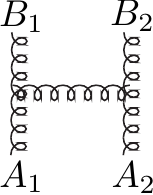} &= C_{\rm H}  \equiv   (- \img f^{A_1 B_1 C} )(-\img f^{A_2 B_2 C})=\frac{N_c}{2}
  \sum_R c_R P_{2\, R}^{A_1 A_2\, B_1 B_2}
  \,.
  \end{align}
Here $P_{2\, R}^{A_1 A_2\, B_1 B_2}$ are color projectors for two gluons, whose explicit expressions can be found in e.g.  \cite{Beneke:2009rj,Keppeler:2012ih}.

However, things get more complicated starting with the three-Glauber exchange case we consider here. 
The color space of 3 gluons in the $t$ channel can be decomposed into irreps as (for $N_c=3$)
\begin{align}\label{eq:888decomp}
  8\otimes8\otimes8=1^2 \oplus 8^8 \oplus 10^4 \oplus \overline{10}^4 \oplus 27^6 \oplus 35^2 \oplus \overline{35}^2 \oplus 64
  \,,
\end{align}
where the number in the superscript denotes the 
multiplicity of $R$'s of a common dimension,
\begin{align}
  R^{N} \equiv R\oplus \cdots \oplus R
  \,.
\end{align}
In this equation the $R$'s on the right hand side are each distinct color representations, but have the same dimension, $\dim(R)$.
The multiplicity $N$ here depends on both the number of gluons $i$ (for the case here, it is 3), and $\dim(R)$. We use a notation $N_{i,R}$ to denote this.
Therefore, for each color channel with  $\dim(R)$, $\gamma_{(i,j)}$ corresponds to a $N_{i,R}\times N_{j,R}$ matrix in color space.

In \sec{structure} we discussed the projection of jet and soft functions onto specific color irreps. 
The jet function $J_{\kappa U(i)}^R$ has a color irrep $U$ for its external scattering state and color irrep $R$ for its $i$ Glauber attachments. The dimension of $U$ and $R$ must agree, $\dim(U)=\dim(R)$, but their multiplicity need not be the same. In color space $J_{\kappa U(i)}^R$ therefore corresponds to a $N_{\kappa,U}\times N_{i,R}$ color matrix, where, $N_{\kappa,U}$ depends on the number of particles and identity of the external state $\kappa$ and the choice of its color irrep $U$. For example, for the case of $\kappa$ being one parton (either gluon or quark), we have $N_{\kappa,U}=1$.\footnote{For $\kappa=g$, according to \eq{2gluon_decomp}, we have $N_{g,U}=N_{2,U}=1$. For $\kappa=q$ or $\bar q$, since $3\otimes\overline{3}=1\oplus 8$ we also have $N_{q,U}=1$. 
}
Then, for each $U$ the jet function $J_{\kappa U(i)}^R$ simply corresponds to a $N_{i,R}\,-\,$dimensional vector.
However, for the case where $\kappa$ contains more particles, e.g., $N_{\kappa=gg,U}$, these jet functions $J_{\kappa U(i)}^R$ also become matrices.

In the following part of this section, by considering the color channels 
that first appear for three-Glauber exchange for NNLL amplitudes, we are able to focus on iterations of the one-loop $\gamma_{(3,3)}$. 
We therefore leave terms involving a more detailed study of transitions of different number of Glaubers to future work (where the simplest case is $\gamma_{(2,3)}$).
These include the odderon (the signature odd singlet in \eq{888decomp}), decupleton $10\oplus\overline{10}$, triantapenton $35\oplus\overline{35}$ and tetrahexaconton $64$. These channels allow us to illustrate the treatment of non-trivial multiplicity of color irreps in the simplest case. Since none of these color channels occur for $8\otimes 8$, they are not affected by $2\to3$ Glauber transitions.

The nontrivial color factors for $\gamma_{(3,3)}$ are the three $C_{{\rm H}_{ij}}$ in \eq{CHij}.
For each irrep, $C_{{\rm H}_{ij}}$ corresponds to $n_{3,R}\times n_{3,R}$ matrices
\begin{align}\label{eq:CH_MH_correspondence}
  C_{{\rm H}_{ij}} \overset{R}{=} M_{{\rm H}_{ij}}^R\,.
\end{align}
which act as transition matrices within each irrep,
while the trivial color factor $\prod_{i=1}^3\delta^{A_iB_i}$ simply correspond to an identity matrix for each $R$.
These matrices can be computed explicitly by decomposition of $C_{{\rm H}_{ij}}$ into the orthogonal 6-gluon basis~\cite{Keppeler:2012ih,Sjodahl:2015qoa} with $\texttt{ColorMath}$ package~\cite{Sjodahl:2012nk}.
Furthermore, the external projection (which is the color factor of the leading order jet function in this case)
corresponds to a vector
\begin{align}\label{eq:color_J_tree}
  J_{\kappa(3)}^{[0]} \overset{R}{=} 
  J_{\kappa(3)}^{R[0]} \equiv g^3 {\cal S}_n V_\kappa^R\,,
\end{align}
where ${\cal S}_n$ are spin factors and $V_\kappa^R$ are color factors. 
The $V_\kappa^R$ can be calculated from Ref.~\cite{Sjodahl:2015qoa} using the 5-gluon base for
$\kappa=g$, or the 2-quark 3-gluon base  for $\kappa=q$.

The outline of the following three subsections is as follows.
In \sec{odderon}, we work out the odderon channel, which can be probed by quark-quark or quark-antiquark scattering.
Since the odderon is one-dimensional (as it can be separated from the Pomeron by its signature), its evolution equation is scalar in terms of color.
In \sec{decuplet}, we consider the interesting decupleton channel which can be probed by gluon-gluon scattering, i.e.~$10\oplus\overline{10}$, where nontrivial multi-dimensional color transitions appear.
In \sec{35_64}, we give the internal color evolutions for $35\oplus\overline{35}$ and $64$. 
Although they are projected to zero for $\kappa$ being single parton $q$ or $g$, one can probe these channels with $\kappa$ and $\kappa'$ that contains more than one parton, e.g., $\kappa,\kappa'=gg$ or $ggg$. We leave a discussion of external projections for these channels to future work.

%%%%%%%%%%%%%%%%%%%%%
\subsection{Odderon}
\label{sec:odderon}
%%%%%%%%%%%%%%%%%%%%%

In the   $8\otimes8\otimes8$ there are two singlet irreps. One is signature even, and is a contribution to the Pomeron, the other is signature odd, and is refferred to as the odderon. The pomeron receives its first contribution from two Glauber exchange, while the odderon first appears with three-Glauber exchange. The evolution equation for the odderon was first computed in \cite{Kwiecinski:1980wb}. Since the odderon has multiplicity 1, it does not yet illustrate the phenomenon of multiple irreps. However, it still provides an illustration of our approach to deriving evolution equations from the rapidity renormalization group, so we include it for completeness.

Since the odderon can be isolated by its signature, each of the three $C_{{\rm H}_{ij}}$'s simply correspond to a number, and they are identical.
\begin{align}\label{eq:CH_odderon}
  C_{{\rm H}_{12}}\,,\, C_{{\rm H}_{23}}\,,\, C_{{\rm H}_{13}} \overset{\text{odderon}}{=} \frac{N_c}{2}\,.
\end{align}
(For the Pomeron, $C_{{\rm H}_{ij}}$ also are all equal to $\frac{N_c}{2}$.)
We then get a differential equation for the odderon which is scalar in color
\begin{align} \label{eq:odderon_RRGE}
  &\nu \frac{\partial}{\partial \nu}J_{q(3)}^{\text{odderon}}(\{\ell_{\perp}\};\nu) \\
       =\,& -\int\! \prod_{i=1}^{3}\frac{\dbar^{d'}\!k_{i\perp}}{k_{i\perp}^2} \,
       J_{q(3)}^{\text{odderon}}(\{k_{\perp}\};\nu)
       \biggl\{ \prod_{i=1}^3\delta^{d'}\!(\ell_{i\perp}- k_{i\perp}) \sum_{i=1}^3\w_G(\ell_{i\perp}) 
       \nn\\
    &\qquad + 
       \delta^{d'}\!(q_\perp-k_{1\perp}-k_{2\perp}-k_{3\perp})\,
       \sum_{i<j}^3 \delta^{d'}(\ell_{i\perp}+\ell_{j\perp}-k_{i\perp}-k_{j\perp})\,  K_\text{NF}(\ell_{i\perp}, \ell_{j\perp};k_{i\perp},k_{j\perp})  \biggr\}
       \,.\nn
 \end{align}

 As a cross check of our evolution equation, we iterate it twice, project to the external quark states and calculate the amplitude at three and four loops. We have confirmed these results agree with App.~D of Ref.~\cite{Falcioni:2021buo}.

%%%%%%%%%%%%%%%%%%%%%%%%%%%%%%%%
\subsection{$10\oplus\overline{10}$}\label{sec:decuplet}
%%%%%%%%%%%%%%%%%%%%%%%%%%%%%%%%

Next we study the evolution for  $10\oplus\overline{10}$, the decuplet channel, for gluon-gluon scattering.
There are 6 copies of decuplets for general $N_c$, so that $\Mdecuplet_{H_{ij}}$ correspond to $6\times6$ matrices. For $N_c=3$, two copies among the six get decoupled and one can only probe four of them, which is why in \eq{888decomp} there are four instead of six copies of decuplets.

In \sec{gen_Nc}, we consider the general $N_c$ case. 
In \sec{Nc_3}, we take $N_c=3$, and see that a 2-d subspace decouples from the 4-d physically accessible subspace.
In \sec{large_Nc}, we take the large-$N_c$ limit, and find a closed form solution to the decupleton evolution equation to all orders.

\subsubsection{General $N_c$}\label{sec:gen_Nc}

The external color of $J_{g(3)}^{[0]}$ can be written as the vector
\begin{align}\label{eq:Vg}
  \Vdecuplet =\frac12\sqrt{\frac{N_c}{N_c^2-4}}
  \begin{pmatrix}
    \begin{array}{cccc:cc}
    N_c\,,&0\,,& \sqrt{(N_c-2)(N_c+3)}\,,&0\,,&\,\,0\,,&-\sqrt{(N_c+2)(N_c-3)}
  \end{array}
  \end{pmatrix}^T
  \,,
\end{align}
and the three transition colors $\Ma$, $\Mb$, $\Mc$ are $6\times6$ matrices.
A nice property for the decupleton channel is that in the basis of Ref.~\cite{Sjodahl:2015qoa} the matrix $\Mc$ is
diagonal, and the other two can be related by an $SO(6)$ rotation $R$,
\begin{align}\label{eq:Mc}
  \Mc &= \text{diag}
  \begin{pmatrix}
    \begin{array}{cccc:cc}
    \frac{N_c}{2}\,,&\frac{N_c}{2}\,,&-1\,,&0\,,&\,\,0\,,&1
  \end{array}
\end{pmatrix} \,, \\[.2cm]
\label{eq:Ma}
\Ma &=R^T\, \Mc\, R\,,
\\[.2cm]
\label{eq:Mb}
\Mb &=R\, \Mc\, R^T\,.
\end{align}
Here, $R$ satisfies the property of $R^3=1$, is found to be\footnote{We note that $R$ has eigenvalues $1,1,e^{+\img 2\pi/3},e^{-\img 2\pi/3},e^{+\img 2\pi/3},e^{-\img 2\pi/3}$.
Therefore, it keeps a 2-dimensional plane invariant, and
corresponds to rotations of the remaining 4-dimensional subspace uniformly by an angle $2\pi/3$, albeit with degenerate eigenvalues.}
\begin{align}\label{eq:R}
  R=\begin{pmatrix}\small
    \begin{array}{cccc:cc}
    \frac{2}{N_c^2 - 4 } & -\frac{1}{\sqrt{N_c^2 - 4 }} & \frac{N_c \sqrt{\frac{N_c+3}{N_c-2}}}{2 (N_c + 2)} & -\frac{1}{\sqrt{2(N_c^2 - 4)}} & -\sqrt{\frac{N_c^2 - 9}{2(N_c^2 - 4) }} & -\frac{N_c \sqrt{\frac{N_c-3}{N_c + 2}}}{2(N_c - 2)}\\[.4cm]
    \frac{1}{\sqrt{N_c^2 - 4 }} & 0 & -\frac{1}{2} \sqrt{\frac{N_c+3}{N_c + 2}} & -\frac{1}{\sqrt{2}} & 0 & -\frac{1}{2} \sqrt{\frac{N_c - 3}{N_c - 2}}\\[.4cm]
    \frac{N_c\sqrt{\frac{N_c + 3}{N_c - 2}}}{2 (N_c + 2)}  & \frac{1}{2} \sqrt{\frac{N_c+3}{N_c + 2}} & \frac{1}{2 (N_c + 2)} & \frac12\sqrt{\frac{N_c+3}{2(N_c + 2)}} & \frac12\sqrt{\frac{N_c - 3}{2(N_c + 2)}} & -\frac{1}{2} \sqrt{\frac{N_c^2 - 9}{N_c^2 - 4 }}\\[.4cm]
    \frac{1}{\sqrt{2} \sqrt{N_c^2 - 4 }} & -\frac{1}{\sqrt{2}} & -\frac12\sqrt{\frac{N_c+3}{2(N_c + 2)}} & \frac{1}{2} & 0 & -\frac12 \sqrt{\frac{N_c-3}{2(N_c-2)}}\\[.4cm]\hdashline \noalign{\vskip 5pt}
    \sqrt{\frac{N_c^2 - 9}{2(N_c^2 - 4) }} & 0 & -\frac12\sqrt{\frac{N_c - 3}{2(N_c + 2)}} & 0 & -\frac{1}{2} & \frac12 \sqrt{\frac{N_c+3}{2(N_c-2)}}\\[.4cm]
    \frac{-N_c \sqrt{\frac{N_c - 3}{N_c + 2}}}{2(N_c-2)} & \frac{1}{2} \sqrt{\frac{N_c - 3}{N_c - 2}} & -\frac{1}{2} \sqrt{\frac{N_c^2 - 9}{N_c^2 - 4 }} & \frac12 \sqrt{\frac{N_c-3}{2(N_c-2)}} & -\frac12 \sqrt{\frac{N_c+3}{2(N_c-2)}} & -\frac{1}{2(N_c-2)}
    \end{array}
\end{pmatrix}\,.
\end{align}
The dashed lines in \eqs{Mc}{R} will be useful for the case of $N_c=3$ in \sec{Nc_3}, since we see that it delineates the off diagonal components which vanish for $N_c=3$, making it easy to see how the two parts of the matrix decouple.

We also notice that the transformation matrix $D=\text{diag}(1,-1,1,-1,-1,1)$ takes $R$ to $R^T=R^2$,
\begin{align}
  DRD=R^T = R^2\,.
\end{align}
In other words, it switches $\Ma$ and $\Mb$ while keeping $\Mc$ invariant,
\begin{align}
 D \Ma D=\Mb\,,\qquad
 D \Mb D=\Ma\,,\qquad
 D \Mc D=\Mc\,.
\end{align}
We further notice that both $R$ and $D$ keep the external vector invariant,
\begin{align}
  R\,\Vdecuplet = \Vdecuplet\,,\qquad
  D\,\Vdecuplet = \Vdecuplet\,.
\end{align}
Therefore, a combination of transformations $R$ and $D$ gives a permutation symmetry in terms of $\bigl\{\Ha,\Hb,\Hc\bigr\}$:
for the color factor of any iteration 
\begin{align}\label{eq:iteration_color}
  \Bigl(\Vdecuplet\Bigr)^T \Mdecuplet_{i_1} \Mdecuplet_{i_2}\cdots \Mdecuplet_{i_n}\, \Vdecuplet\,,
\end{align}
where $i_1,\cdots ,i_n\in \{\Ha,\Hb,\Hc\}$,
switching all $\Ma \leftrightarrow \Mb$ or $\Ma \leftrightarrow \Mc$ or $\Mb\leftrightarrow \Mc$ does not change the result. Therefore, for any color factor as in \eq{iteration_color}, 
these swaps will not change the result. 

The differential equation for the decoupletons can be written in terms of  these color matrices and is
\begin{align} \label{eq:10_RRGE}
  &\nu \frac{\partial}{\partial \nu}J_{g(3)}^{10\oplus\overline{10}}(\{\ell_{\perp}\};\nu) \\
       &=\, -\int\! \prod_{i=1}^{3}\frac{\dbar^{d'}\!k_{i\perp}}{k_{i\perp}^2} \,
       J_{g(3)}^{10\oplus\overline{10}}(\{k_{\perp}\};\nu)
       \biggl\{ \prod_{i=1}^3\delta^{d'}\!(\ell_{i\perp}- k_{i\perp}) \sum_{i=1}^3\w_G(\ell_{i\perp}) 
       \nn\\
    &\ \ + \frac{2}{N_c}
       \delta^{d'}\!(q_\perp-k_{1\perp}-k_{2\perp}-k_{3\perp})\,
       \sum_{i<j}^3 \delta^{d'}(\ell_{i\perp}+\ell_{j\perp}-k_{i\perp}-k_{j\perp})\,\Mdecuplet_{\text{H}_{ij}}\,  K_\text{NF}(\ell_{i\perp}, \ell_{j\perp};k_{i\perp},k_{j\perp})  \biggr\}
       .\nn
 \end{align}
It takes the form of a BFKL equation, but in the 6$\times$6 space. This is a primary result of this paper, and illustrates how we can describe evolution equations of representations with non-trivial multiplicity. 

As a check of our evolution equation, we have iterated it once and twice to three and four loops and then projected with $\kappa\kappa'=gg$ which collapses it to a single result at each order in $\alpha_s$.  We have confirmed these results agree perfectly with App.~D\footnote{When making the comparison, we note that their expressions are given for so-called reduced amplitudes by pulling out a factor of $\exp\bigl(-\mathbf{T}_t^2 \alpha_g(t)L\bigr)$, whereas our RGE iterates to the full amplitudes. This subtlety does not matter for the odderon check in the previous section, given that $\mathbf{T}_t^2=0$ for color singlets; but it matters here for $10\oplus\overline{10}$.} of Ref.~\cite{Falcioni:2021buo}, providing a nontrivial cross check on our more general matrix equation.

%%%%%%%%%%%%%%%%%%
\subsubsection{$N_c=3$}
\label{sec:Nc_3}
%%%%%%%%%%%%%%%%%%
For $N_c=3$, we notice that $R$ in \eq{R} becomes block diagonal in the way the dashed lines indicate. 
Also notice that the last entry of \eq{Vg} is zero.
Therefore, the two-dimensional space below (or at the RHS of) the dashed lines will never be probed, and 
it decouple from the 4-dimensional space.
We can keep only the first block of the vector, and the matrices, which we explicitly write down,
\begin{align}\label{eq:Vg3}
  V_g^{10\oplus\overline{10}}=\frac{\sqrt{15}}{10}
  \begin{pmatrix}
    \begin{array}{cccc}
    3\,,&0\,,& \sqrt{6}\,,&0
  \end{array}
  \end{pmatrix}^T
  \,,
\end{align}
\begin{align}\Ma &=
\begin{pmatrix}
  0 & -\frac{3\sqrt{5}}{10} & 0 & -\frac{3\sqrt{10}}{10} \\
  -\frac{3\sqrt{5}}{10} & 0 & -\frac{\sqrt{30}}{10} & 0 \\
  0 & -\frac{\sqrt{30}}{10} & \frac{5}{4} & \frac{\sqrt{15}}{20} \\
  -\frac{3\sqrt{10}}{10} & 0 & \frac{\sqrt{15}}{20} & \frac{3}{4}
  \end{pmatrix}\,,
  \quad
  \Mb =
  \begin{pmatrix}
    0 & \frac{3\sqrt{5}}{10} & 0 & \frac{3\sqrt{10}}{10} \\
    \frac{3\sqrt{5}}{10} & 0 & \frac{\sqrt{30}}{10} & 0 \\
    0 & \frac{\sqrt{30}}{10} & \frac{5}{4} & -\frac{\sqrt{15}}{20} \\
    \frac{3\sqrt{10}}{10} & 0 & -\frac{\sqrt{15}}{20} & \frac{3}{4}
    \end{pmatrix}\,,
  \quad
  \nn\\[.2in]
  \Mc &=
  \begin{pmatrix}
    3/2 & 0 & 0 & 0 \\
    0 & 3/2 & 0 & 0 \\
    0 & 0 & -1 & 0 \\
    0 & 0 & 0 & 0
  \end{pmatrix}
  \,,
  \quad
  R=
  \begin{pmatrix}
    \frac{2}{5} & -\frac{\sqrt{5}}{5} & \frac{\sqrt{6}}{10} & -\frac{\sqrt{10}}{10} \\[.15cm]
    \frac{\sqrt{5}}{5} & 0 & -\frac{\sqrt{30}}{10} & -\frac{\sqrt{2}}{2} \\[.15cm]
    \frac{\sqrt{6}}{10} & \frac{\sqrt{30}}{10} & \frac{1}{10} & \frac{\sqrt{15}}{10} \\[.15cm]
    \frac{\sqrt{10}}{10} & -\frac{\sqrt{2}}{2} & -\frac{\sqrt{15}}{10} & \frac{1}{2}
\end{pmatrix}
\,.
\end{align}
$R$ here has eigenvalues to be $1,1,e^{\img 2\pi/3},e^{-\img 2\pi/3}$. So it keeps a two-dimensional plane invariant and rotate an orthogonal two-dimensional plan by an angle $2\pi/3$.

\subsubsection{Large $N_c$ limit}\label{sec:large_Nc}
Keeping the leading term in the $N_c\to\infty$ limit, we note that
\begin{align}
  \Mc=\frac{N_c}{2}\,\text{diag}(1,1,0,0,0,0)\,,
\end{align}
and
\begin{align}
  \Mdecuplet_i\Mdecuplet_j =0\,,\quad\text{for $i\neq j$}\,.
\end{align}
Therefore, 
by simply noticing that $\Ma$ and $\Mb$ can be diagonalized to $\Mc$, a base can be found such that
\begin{align}
  &\Ma=\frac{N_c}{2}\,\text{diag}
  \begin{array}{ccc:ccc}
    \bigl(1\,,&0\,,& 0\,,&\,1\,,&0\,,&0\bigr)
  \end{array}
  \,,
  \qquad
  \Mb=\frac{N_c}{2}\,\text{diag}
  \begin{array}{ccc:ccc}
    \bigl(0\,,&1\,,& 0\,,&\,0\,,&1\,,&0\bigr)
  \end{array}
  \,,
  \\[.2cm]\nn
  &\Mc=\frac{N_c}{2}\,\text{diag}
  \begin{array}{ccc:ccc}
    \bigl(0\,,&0\,,& 1\,,&\,0\,,&0\,,&1\bigr)
  \end{array}
  \,,
  \\[.2cm]
  &
  \Vdecuplet =\frac12 \sqrt{N_c}
    \begin{array}{ccc:ccc}
      \bigl(1\,,&1\,,& 1\,,&\,0\,,&0\,,&0\bigr)
    \end{array}^{\!\!T}
  \,.
\end{align}  
In this base, the 3-dimensional space RHS of the dashed lines are not relevant, and we simply ignore them. We find a simple solution
\begin{align}
  J_{g(3)}^{10\oplus\overline{10}}(\ell_{1\perp},\ell_{2\perp},\ell_{3\perp};\nu)\sim \left\{\left(\frac{s}{\nu^2}\right)^{
\w_G(q_\perp-\ell_{1\perp})+\w_G(\ell_{1\perp})},
\left(\frac{s}{\nu^2}\right)^{
\w_G(q_\perp-\ell_{2\perp})+\w_G(\ell_{2\perp})},
\left(\frac{s}{\nu^2}\right)^{
\w_G(q_\perp-\ell_{3\perp})+\w_G(\ell_{3\perp})}
\right\}\,,
\end{align}
and the amplitude can be resummed to a closed form solution,
\begin{align}
  \cM^{10\oplus\overline{10}} &\propto
  \int\!\frac{\dbar^{d'}\!\ell_{1\perp}\dbar^{d'}\!\ell_{2\perp} \dbar^{d'}\!\ell_{3\perp}}{\vec\ell_{1\perp}^{\,2}\vec\ell_{2\perp}^{\,2} \vec\ell_{3\perp}^{\,2}}\,
  \deltaslash^{d'}\!\bigl(\ell_{1\perp}+\ell_{2\perp}+\ell_{3\perp}-q_\perp\bigr)\,
  \sum_{i=1}^3
\left(\frac{s}{-t}\right)^{
\w_G(q_\perp-\ell_{i\perp})+\w_G(\ell_{i\perp})}
\nn\\[.2cm]
&=
\frac{3B(1,1)}{(4\pi)^{1-\epsilon}}
\int\!\frac{\dbar^{d'}\!\ell_{1\perp}}{\vec\ell_{1\perp}^{\,2} \bigl(\vec q_\perp-\vec\ell_{1\perp}\bigr)^{\!2+2\epsilon}}
\left(\frac{s}{-t}\right)^{
\w_G(q_\perp-\ell_{1\perp})+\w_G(\ell_{1\perp})}
\,,
\end{align}
where 
\begin{align} \label{eq:Bab}
    B(\alpha, \beta) \equiv \frac{\Gma{1-\alpha-\epsilon} \Gma{1-\beta-\epsilon} \Gamma(\alpha+\beta-1+\epsilon)}{
    \Gma{\alpha} \Gma{\beta} \Gma{2-\alpha-\beta -2\epsilon}
    } . 
\end{align}

%%%%%%%%%%%%%%%%%%%%%%%%%%%%%%%%%%
\subsection{$35\oplus\overline{35}$ and 64}
\label{sec:35_64}
%%%%%%%%%%%%%%%%%%%%%%%%%%%%%%%%%%

As further illustrations of our formalism, we can also consider the $35\oplus\overline{35}$ and $64$ representations. These are not usually considered, since they cannot be excited in the $2\to 2$ forward scattering. However, they can be excited in more general scattering processes, so we consider their evolution for completeness.

The multiplicity of the $35\oplus\overline{35}$ irrep in three-Glauber exchange is 2. It therefore behaves similarly to the decuplet channel considered previously. The three matrices as
\begin{align}\label{eq:M35}
  \Mcc&=\text{diag}(0,-1)\,,\\[.2cm]
  \Maa&=R_{2\pi/3}^T\,\Mcc\, R_{2\pi/3}\,,\\[.2cm]
  \Mbb &=R_{2\pi/3}\,\Mcc\, R_{2\pi/3}^T
  \,,
\end{align}
where $R_{2\pi/3}$ is a 2-dimensional rotation with angle $2\pi/3$
\begin{align}
  R_{2\pi/3}=
  \begin{pmatrix}
    -\frac12 & \frac{\sqrt{3}}{2} \\[.15cm]
    -\frac{\sqrt{3}}{2} & -\frac12
\end{pmatrix}\,.
\end{align}
Also, we find that there exists a transformation $\text{diag}(1,-1)$ that takes $R_{2\pi/3}$ to $R_{2\pi/3}^T$, and switches $\Maa$ with $\Mbb$ while keeping $\Mcc$ invariant.
The resulting evolution equation for triantapentons is
\begin{align} \label{eq:35_RRGE}
  &\nu \frac{\partial}{\partial \nu}J_{\kappa(3)}^{\text{35}\oplus\overline{\text{35}}}(\{\ell_{\perp}\};\nu) \\
       =\,& -\int\! \prod_{i=1}^{3}\frac{\dbar^{d'}\!k_{i\perp}}{k_{i\perp}^2} \,
       J_{\kappa(3)}^{\text{35}\oplus\overline{\text{35}}}(\{k_{\perp}\};\nu)
       \biggl\{ \prod_{i=1}^3\delta^{d'}\!(\ell_{i\perp}- k_{i\perp}) \sum_{i=1}^3\w_G(\ell_{i\perp}) 
       \nn\\
    &\ \ + \frac{2}{N_c} 
       \delta^{d'}\!(q_\perp-k_{1\perp}-k_{2\perp}-k_{3\perp})\,
       \sum_{i<j}^3 \delta^{d'}(\ell_{i\perp}+\ell_{j\perp}-k_{i\perp}-k_{j\perp})\, M^{35\oplus\overline{35}}_{\text{H}_{ij}} \,K_\text{NF}(\ell_{i\perp}, \ell_{j\perp};k_{i\perp},k_{j\perp})  \biggr\}
       \,.\nn
 \end{align}
 As for the $10\oplus \bar{10}$, it takes a BFKL form, this time with a 2$\times$2 matrix. 

The multiplicity for the $64$ irrep is $1$. Therefore, $C_{\text{H}_{ij}}$ corresponds to a number. We find that they are identical for the three H$_{ij}$'s,
\begin{align}\label{eq:M64}
  M_{\text{H}_{ij}}^{64}=-1
  \,.
\end{align}
The resulting evolution equation for tetrahexaconton is
\begin{align} \label{eq:64_RRGE}
  &\nu \frac{\partial}{\partial \nu}J_{\kappa(3)}^{64}(\{\ell_{\perp}\};\nu) \\
       =\,& -\int\! \prod_{i=1}^{3}\frac{\dbar^{d'}\!k_{i\perp}}{k_{i\perp}^2} \,
       J_{\kappa(3)}^{64}(\{k_{\perp}\};\nu)
       \biggl\{ \prod_{i=1}^3\delta^{d'}\!(\ell_{i\perp}- k_{i\perp}) \sum_{i=1}^3\w_G(\ell_{i\perp}) 
       \nn\\
    &\qquad - 
    \frac{2}{N_c}\,\delta^{d'}\!(q_\perp-k_{1\perp}-k_{2\perp}-k_{3\perp})
       \sum_{i<j}^3 \delta^{d'}(\ell_{i\perp}+\ell_{j\perp}-k_{i\perp}-k_{j\perp})\,  K_\text{NF}(\ell_{i\perp}, \ell_{j\perp};k_{i\perp},k_{j\perp})  \biggr\}
       \,.\nn
 \end{align}

Interestingly, we note that for both $R=35\oplus\overline{35}$ and $R=64$,
$M_{\text{H}_{ij}}^{R}$ are
$1/N_c$ suppressed compared to the Regge pole terms in $\gamma_{(3,3)}$, i.e., we only need to keep the green blobs in \eq{gamma33_graph}. Therefore, in the large $N_c$ limit, for both $R=35\oplus\overline{35}$ and $64$, it easily resums to
\begin{align}\label{eq:35_64}
  \cM_{K\to K}^{\kappa\kappa'\,R}\propto %\cM_{K\to K}^{[0]\kappa\kappa'\,R} 
  \int\!\frac{\dbar^{d'}\!\ell_{1\perp}\dbar^{d'}\!\ell_{2\perp}}{\vec\ell_{1\perp}^{\,2}\vec\ell_{2\perp}^{\,2} \vec\ell_{3\perp}^{\,2}} \biggl(\frac{s}{-t}\biggr)^{\w_G(\ell_{1\perp})+\w_G(\ell_{2\perp})+\w_G(\ell_{3\perp})} \,,
\end{align}
where $\ell_{3\perp}$ is understood to be $q_\perp-\ell_{1\perp}-\ell_{2\perp}$.
It would be interesting to find some explicit external projection to study these channels in physical scattering.

%%%%%%%%%%%%%%%%%%%%%%%%%%%%%%%%%%%%%%%%%%%%%%%%%%%%%%%%%%%%%%%%
\section{Conclusions}
\label{sec:conc}
%%%%%%%%%%%%%%%%%%%%%%%%%%%%%%%%%%%%%%%%%%%%%%%%%%%%%%%%%%%%%%%%

As compared to the simplicity of Regge scattering in the planar limit, despite extensive study, Regge scattering beyond the planar limit remains complicated and its all order structure remains poorly understood. In this paper we have used our recently developed approach to organize forward scattering in terms of Glauber operators \cite{Gao:2024qsg} in the EFT of forward scattering \cite{Rothstein:2016bsq} to elucidate one aspect of this puzzle, namely the treatment of irreps with color multiplicity greater than one. Instead of the standard approach of projecting onto irreps using the external particles, we showed that it is advantageous to first perform internal color projections to derive evolution equations for different irreps that are universal with respect to the scattered particles, but which incorporate irrep multiplicity. The equations take the form of BFKL-type equations, but have an additional matrix structure taking into account the multiplicity of the irrep, and account for mixing in the internal color space.

We illustrated this approach by deriving evolution equations for specific irreps in three-Glauber exchange. To isolate this particular aspect, we focused only on irreps that first appear at three-Glauber level. This prevented us from having to consider the mixing between Regge cut and Regge pole contributions. These irreps are the odderon,  the decoupletons ($10\oplus\overline{10}$), the triantepentons ($35\oplus\overline{35}$) and the tetrahexaconton ($64$). We derived evolution equations for each of these irreps by performing the rapidity renormalization of the collinear sector, using the approach developed in \cite{Gao:2024qsg}. For the decoupletons and the triantepentons, these equations involve a non-trivial matrix structure due to the non-trivial multiplicity of these irreps. To our knowledge, the treatment of irreps with higher multiplicity has not been treated previously in the small-$x$ or Regge literature, and these examples show how it can be systematically treated in the EFT for forward scattering.

In a forthcoming paper, we will extend our approach to study the remaining channels, namely the Pomeron, the $8$, and the $27$, which have the added complexity of involving a mixing with operators involving different numbers of Glauber operators. Our approach also provides a clean separation of Regge cut and Regge pole contributions, and will further illustrate how we can systematically organize Regge scattering beyond the planar limit within the EFT approach. It will be extremely interesting to compare the result for the three-loop Regge trajectory computed in this approach with that computed using the Reggeon Wilson-line Hamiltonian based organizing principles used in Refs.~\cite{Caola:2021izf,Falcioni:2021dgr}.

Although this paper has been focused on the Regge limit of $K\to K$ scattering with a single large rapidity gap, and thus a single type of large logarithm, our general organization applies also to the multi-Regge limit. Much like the case considered in this paper, multi-Regge scattering is well understood in planar theories, particularly $\mathcal{N}=4$ super Yang-Mills \cite{Kuraev:1976ge,Fadin:2006bj,Bartels:2008ce,Bartels:2008sc,Dixon:2012yy,Basso:2014pla,Bargheer:2015djt,DelDuca:2016lad,Bargheer:2016eyp,Bargheer:2016eyp,DelDuca:2017peo,DelDuca:2018raq,DelDuca:2018hrv}, but poorly understood beyond the planar limit. Our general approach based on the EFT of forward scattering is well posed to address these interesting questions.

%%%%%%%%%%%%%%%%%%%%%%%%%%%%%%%%%%%%%%%%%%%%%%%%%%%%%%%%%%%%%%%%
\begin{acknowledgments}
We thank Ira Rothstein and Michael Saavedra for useful discussions.  This work was supported in part by the  U.S. Department of Energy, Office of Nuclear Physics from DE-SC0011090 and the Simons Foundation through the Investigator grant 327942.
We thank the Erwin Schr\"odinger Institute where part of this work was completed during the 2023 program on Quantum Field Theory at the Frontiers of the Strong Interaction.
\end{acknowledgments}
%%%%%%%%%%%%%%%%%%%%%%%%%%%%%%%%%%%%%%%%%%%%%%%%%%%%%%%%%%%%%%%%
  
\bibliography{../reggeNLL_bib}{}
\bibliographystyle{JHEP}

\end{document}